\documentclass[11pt]{article}

\usepackage{fullpage}
\usepackage{multirow}

\usepackage[utf8]{inputenc}
\usepackage{titlesec}
\usepackage{graphicx}
\usepackage{tabularx}
\usepackage{bm}
\usepackage{caption}
\usepackage{color}
\usepackage{amsmath}
\usepackage{float}
\usepackage{bm}
\usepackage{setspace}
\usepackage{chngcntr}
\usepackage[page,title]{appendix}
\usepackage{enumerate}% http://ctan.org/pkg/enumerate
\usepackage{hyperref}
\hypersetup{
    colorlinks=true,
    linkcolor=blue,
    citecolor=blue,
}
\usepackage{natbib}
\titleformat{\section}[block]{\Large\bfseries\filcenter}{\thesection}{1em}{}
\titleformat{\subsection}[block]{\Large\itshape\filcenter}{\thesubsection}{1em}{}
\titleformat{\subsubsection}[block]{\large\itshape}{\thesubsubsection}{1em}{}
\titleformat{\paragraph}[runin]{\itshape}{\theparagraph}{1em}{}[. ]

\title{How range residency and long-range perception change encounter rates}

\author{Ricardo Martinez-Garcia$^{1}$, Christen H. Fleming$^{2,3}$, Ralf Seppelt$^{4,5}$,\\
William F. Fagan$^{3}$, Justin M. Calabrese$^{2,3,\ast}$}
 
\date{\vspace{-5ex}}
 
\begin{document}
 \maketitle
\begin{center}
\noindent{}1. Dept. of Ecology \& Evolutionary Biology, Princeton University, Princeton NJ 08544, USA. \\
\noindent{}2. Smithsonian Conservational Biology Institute, National Zoological Park, 1500 Remount Road, Front Royal, Virginia 22630 USA. \\
\noindent{}3. Dept. of Biology, University of Maryland, College Park MD 20742, USA. \\
\noindent{}4. Dept. of Computational Landscape Ecology, Helmholtz Centre for Environmental Research–UFZ, Leipzig, Germany. \\
\noindent{}5. Institute of Geoscience \& Geography, Martin-Luther-University Halle-Wittenberg, Halle (Saale), Germany.\\
 \end{center}
\noindent{} $\ast$ Corresponding author; e-mail: calabresej@si.edu. 

\section*{Abstract}
Encounter rates link movement strategies to intra- and inter-specific interactions, and therefore translate individual movement behavior into higher-level ecological processes. Indeed, a large body of interacting population theory rests on the law of mass action, which can be derived from assumptions of Brownian motion in an enclosed container with exclusively local perception. These assumptions imply completely uniform space use, individual home ranges equivalent to the population range, and encounter dependent on movement paths actually crossing. Mounting empirical evidence, however, suggests that animals use space non-uniformly, occupy home ranges substantially smaller than the population range, and are often capable of nonlocal perception. Here, we explore how these empirically supported behaviors change pairwise encounter rates. Specifically, we derive novel analytical expressions for encounter rates under Ornstein-Uhlenbeck motion, which features non-uniform space use and allows individual home ranges to differ from the population range. We compare OU-based encounter predictions to those of Reflected Brownian Motion, from which the law of mass action can be derived. For both models, we further explore how the interplay between the scale of perception and home range size affects encounter rates. We find that neglecting realistic movement and perceptual behaviors can systematically bias encounter rate predictions. 
 
\section*{Introduction}

A key goal of movement ecology is to use knowledge of movement behaviors to understand large-scale ecological processes \citep{nathan2008movement,Mueller2008,Hawkes2009,Morales2010,Atkins2019}. This `upscaling' from individual movement to population- and community-level consequences, including competition, predation, mate finding, and disease transmission, will be mediated by pairwise interactions. A key step in the transition from individual movement to higher-level dynamics is to understand how movement behaviors translate into encounter rates among potentially interacting individuals. Indeed, encounter rates can be seen as a key contact point in movement ecology, as they directly govern how different movement strategies affect intra- and interspecific interactions \citep{CSHolling1959,Turchin2003,Merrill2010,Barraquand2013}.

Analytical results on encounter rates are typically derived from two extremes of a continuum of movement processes with intermediate models being explored through numerical simulations \citep{Blable2016}. On the one end is simple ballistic motion,  which describes straight-line movement of individuals in random directions \citep{Mosimann1958,Gerritsen1977,Hutchinson2007}. On the other extreme is Brownian motion (BM), which describes infinitely tortuous trajectories \citep{Visser2006,Visser2008}. Both types of movement, when confined to an enclosed container with reflecting boundaries, and when coupled with purely local perception, lead to the frequently invoked law of mass action. Mass-action encounter has been the cornerstone of interacting population models going back to the Lotka-Volterra predator-prey equations \citep{Lotka1926,Volterra1926}, and the Kermack-McKendrick SIR disease transmission model \citep{kermack1927contribution}. More recently, mass-action encounters have been also applied to investigate sexual reproduction in single species models \citep{snyder2017mass}. Therefore, a huge swath of ecological theory now rests on the mass action assumption.

For much of the history of ecology, elemental movement processes such as ballistic and Brownian motion have been used to understand the consequences of animal movement. Key reasons for the ubiquity of these models include their analytical tractability, and the long-standing dearth of empirical data with which to better characterize real movement processes. Recent advances in animal tracking technology \citep{Cagnacci2010,coyne2005satellite,kays2015terrestrial}, movement modeling \citep{Patterson2008,Gurarie2011,Fleming2014,Fleming2014b,Pyke2015}, and statistical analyses \citep{kie2010home,Fleming2015,hooten2017animal,winner2018statistical} have removed both of these bottlenecks, and we are now in position to critically reassess the core assumptions undergirding classical encounter theory. Two immediately apparent areas where existing encounter models fall short of empirical reality are in adequately accounting for range residency and nonlocal perception.

In the former case, ecologists have long known that most animals move within well-defined home ranges, and a large and diverse literature has developed around how to best estimate home ranges from tracking data \citep{Odum1955,Jennrich1969,Dixon1980,Worton1987,Worton1989,Powell2000, Fleming2015}. Mounting empirical evidence suggests that animals tend to use their home ranges unevenly and that individual home ranges typically cover only a fraction of the population range \citep{Burt1943,Bowen1982,harris1990home,Fieberg2005,kie2010home,moorcroft2013mechanistic,Benson2015,winner2018statistical}. In stark contrast, simple bounded movement processes that lead to mass action, like Reflected Brownian Motion \citep[RBM;][]{Harrison1985, dieker2010reflected}, result in uniform space use and individual ``home ranges'' that are equivalent to the population range. Similarly, increasing evidence now suggests an important and widespread role for non-local perception in shaping the movement and foraging strategies of many animals \citep{zollner1997landscape, zollner2000comparing, mech2002using, Prevedello2010, fletcher2013signal,schumacher2017sensory, aben2018call,rios2019levy}, whereas mass-action encounter assumes purely local perception. Even though a finite scale of perception has been considered a tunable parameter in previous studies, its effect on encounter processes have been analyzed mostly through numerical simulations \citep{Bartumeus2002,Martinez-Garcia2013,Martinez-Garcia2014,Martinez-Garcia2017,Fagan2017}, with only very few analytical results existing for one-dimensional dynamics \citep{Bartumeus2014} or very particular spatial distributions of targets \citep{Gurarie2012}.

The Ornstein-Uhlenbeck process \citep[OU;][]{Uhlenbeck1930}, which is related to RBM, is the simplest stochastic movement model that captures non-even space use within a home range, and allows individual home ranges to differ from the population range. To the best of our knowledge, however, no encounter theory exists for this model. Importantly, the OU process is increasingly used in empirical tracking studies, and has well developed statistical estimators \citep{Hines2005,Fieberg2007,Fleming2014b,fleming2017kalman}. It can also serve as the basis for more complicated estimation procedures, including composite movement models \citep{Blackwell1997,Breed2017}, and autocorrelated kernel density estimation (AKDE) \citep{Fleming2015, Fleming2017}. Indeed, in a recent comparative study of home range estimators, the OU process was selected as the AIC-best model on which to base AKDE home range estimates for 128 of 369 datasets \citep{Noonan2019}.

Here, we compare the OU and RBM models to explore how individually restricted movement and uneven space use within home ranges affect encounter processes. For both models, we account for nonlocal perceptual ranges and study how a tunable scale for individual perception interacts with the spatial extent of home ranges to determine encounter frequency. We derive novel, exact expressions for the mean encounter rate of OU and RBM processes to quantify how empirically supported departures from classical encounter assumptions change encounter rates, and supplement these with numerical results where necessary. We outline the conditions under which RBM--and by extension, the law of mass action--fails to provide a reasonable approximation to the more realistic OU process, and discuss the potential consequences of this incongruity for interacting population models that invoke mass action encounter.

\section*{Methods}\label{sec:methods}

\subsection*{Encounter metrics}\label{sec:enc}

For simplicity, we will limit our analysis to the encounter between a pair of individuals in which one of them acts as a searcher and the other as a mobile target. This scenario mimics, for instance, a simple prey-predator encounter. Because we do not consider cross-correlations between the trajectories of the individuals, long-range chasing or avoidance mechanisms are neglected. In consequence, we will introduce and discuss our results in terms of a predator-prey interaction, but generalizations to broader encounter scenarios in which two individuals show home-ranging behavior and long-range perception are straightforward. Throughout the manuscript, we will use a notation in which the subscript $1$ refers to the predator and $2$ to the prey.

In this context, the probability that an encounter, $E$, occurs during some small time interval $dt$ is,
\begin{equation} \label{enc-prob}
 \text{Pr}(E \ \text{in} \ \lbrace t, \ t+dt \rbrace) = \mathcal{E}(r(t))dt
\end{equation}
where $\mathcal{E}(t)$ is the \textit{instantaneous encounter rate} and has units of inverse of time. It captures the likelihood that an encounter might take place and it is therefore a function of the distance between the pair of individuals, $r(t)$. More specifically,
\begin{equation}\label{enc-r}
 \mathcal{E}(r(t)) = \gamma\Phi_q(r(t)),
\end{equation}
where $\gamma$ quantifies the area within the detection range explored by the predator per unit of time and $\Phi_q$ is the encounter kernel. Therefore, $\gamma$ has units of area $\times$ time$^{-1}$ and provides the time$^{-1}$ units to encounter rate $\mathcal{E}$. From a mathematical point of view, the encounter kernel, $\Phi_q$, captures the potential for an encounter to take place provided that individuals are at a distance $r$ from each other. From the biological point of view, the encounter kernel is linked to predator's perception, representing its ability to detect prey that are at a given distance. We assume that predator's attention is distributed over the coverage area of the perceptual range. Hence, for a fixed $\gamma$, its ability to detect prey per unit area decays as the perceptual range increases and its attention is spread more thinly over a larger area. Mathematically, this translates into a normalization of the kernel, $\int \Phi_q(r)dr=1$ \citep{Gurarie2012}. In the following sections, we will use a top-hat encounter kernel defined by the predator perceptual range, $q$,
\begin{equation}\label{enc-kernel}
 \Phi_q(r(t)) = \left\{
	       \begin{array}{ll}
		 0      & \mathrm{if\ } r(t) \ > \ q \\
		 \frac{1}{\pi q^2} & \mathrm{if\ } r(t) \ \le \ q \\
	       \end{array}
	     \right.
\end{equation}
The effect of smoother decays of the encounter kernel with distance is discussed at length in Appendix \ref{app:encounter-kernel} and has also been addressed in the literature \citep{Gurarie2012}.

Finally, because the movement models used to describe the trajectories of both individuals are stochastic, both the distance between individuals and the instantaneous encounter rate are stochastic processes. We introduce the mean encounter rate, defined as the instantaneous encounter rate averaged over realizations of the movement processes,
\begin{equation} \label{eq:mean-encounter}
 \tilde{\mathcal{E}}(t) = \int \mathcal{E}(r(t)) f(r(t))dr,
\end{equation}
where $f(r(t))$ is the probability density function (PDF) of the time-dependent distance between individuals, $r(t)$, and $\mathcal{E}(r(t))$ is the instantaneous encounter rate as defined in Eq.~(\ref{enc-prob}).

\subsection*{Movement models}\label{sec:OU}
To investigate the influence of home ranging on encounter processes, we compare OU-based encounter statistics with those derived from RBM. OU allows individual home ranges to differ from the population range and also permits a non-uniform use of the space within each home range. RBM, in contrast, leads to individual home ranges that are equal to the range of the population and to individuals exploring the entire population range uniformly. As an additional point of contrast, we supplement our analysis with encounter statistics derived for Brownian motion (BM), which allows us to investigate the long-term impact that unrestricted movement has on encounter rates (details of the BM calculations are provided in Appendix \ref{app-BM}).

\subsubsection*{Ornstein-Uhlenbeck}
In two dimensions, the OU movement model is described by a pair of independent stochastic differential equations,
\begin{equation}
  \dot{z}_\beta(t)= -\frac{1}{\tau}\left[z_{\beta}(t) - \lambda_\beta \right] + \sqrt{g} \xi_\beta(t), \label{ou-pos}\\
\end{equation}
where the subscript $\beta$ indicates each of the two coordinates in the $2D$ space: $\beta\in\lbrace x,y\rbrace$. $z$ is the location of the individual and the dot indicates a time derivative. $(\lambda_x,\lambda_y)$ gives the home range center \citep{Okubo2001a}, and $(\xi_x,\xi_y)$ is a zero-mean and unit-variance Gaussian white noise process with units of time$^{-1/2}$. Note that the OU model neglects correlations in the velocity and hence the model equations only describe changes in individual position.  For simplicity, we impose isotropy on the movement, so $\tau$ and $g$ are scalar quantities. $\tau^{-1}$ accounts for the average home-range crossing rate, has units of time$^{-1}$, and quantifies the rate at which individuals return to the center of their home ranges after a stochastically initiated excursion that takes them away from it. Therefore, $\tau$ provides a metric for home-range affinity, from nomadic species at $\tau^{-1}\rightarrow0$ to sedentary species that do not abandon the home-range center when $\tau^{-1}\rightarrow\infty$. Finally, $g$ sets the size of the home range by modulating the intensity of the stochastic component of the movement and has units of area/time.

Because the only stochastic terms in Eq.~(\ref{ou-pos}), $\xi_\beta$, are Gaussian processes, the OU model is also a Gaussian process, and is therefore completely defined by the mean of each coordinate, $\mu_\beta$, and the variance $\sigma^2$ (see Appendix \ref{app-ou} for detailed calculations),
\begin{eqnarray}
 \mu_\beta(t) &=& \mu_\beta(0){\rm e}^{-t/\tau} + \lambda_\beta\left( 1-{\rm e}^{-t/\tau} \right), \label{mean-pos} \\
 \sigma^2(t) &=& \sigma^2(0){\rm e}^{-2t/\tau} + \frac{g\tau}{2}\left(1- {\rm e}^{-2t/\tau} \right), \label{var-pos}
\end{eqnarray}
where we have considered that the initial condition is likely to be stochastic and have thus maintained $\mu_\beta(0)$ and $\sigma^2(0)$ in Eqs.~(\ref{mean-pos}) and (\ref{var-pos}), respectively. For deterministic initial conditions, $\sigma^2(0)=0$ and ${\bm \mu}(0)$ is the initial position of the individual. Notice that, because we assume isotropy in the movement, the variance $\sigma^2(t)$ is the same for each component of the position and we do not include the subscript $\beta$ in Eq.~(\ref{var-pos}). In the long-time limit, OU positions converge to a bivariate Normal distribution with mean $(\lambda_x,\lambda_y)$ and variance $g\tau/2$ regardless of the initial conditions. The normality of the positions reflects non-homogeneous individual space use, while the finite variance leads to home ranges that can be smaller than the range of the population.

\begin{figure}
    \centering
        \includegraphics[width=0.5\textwidth]{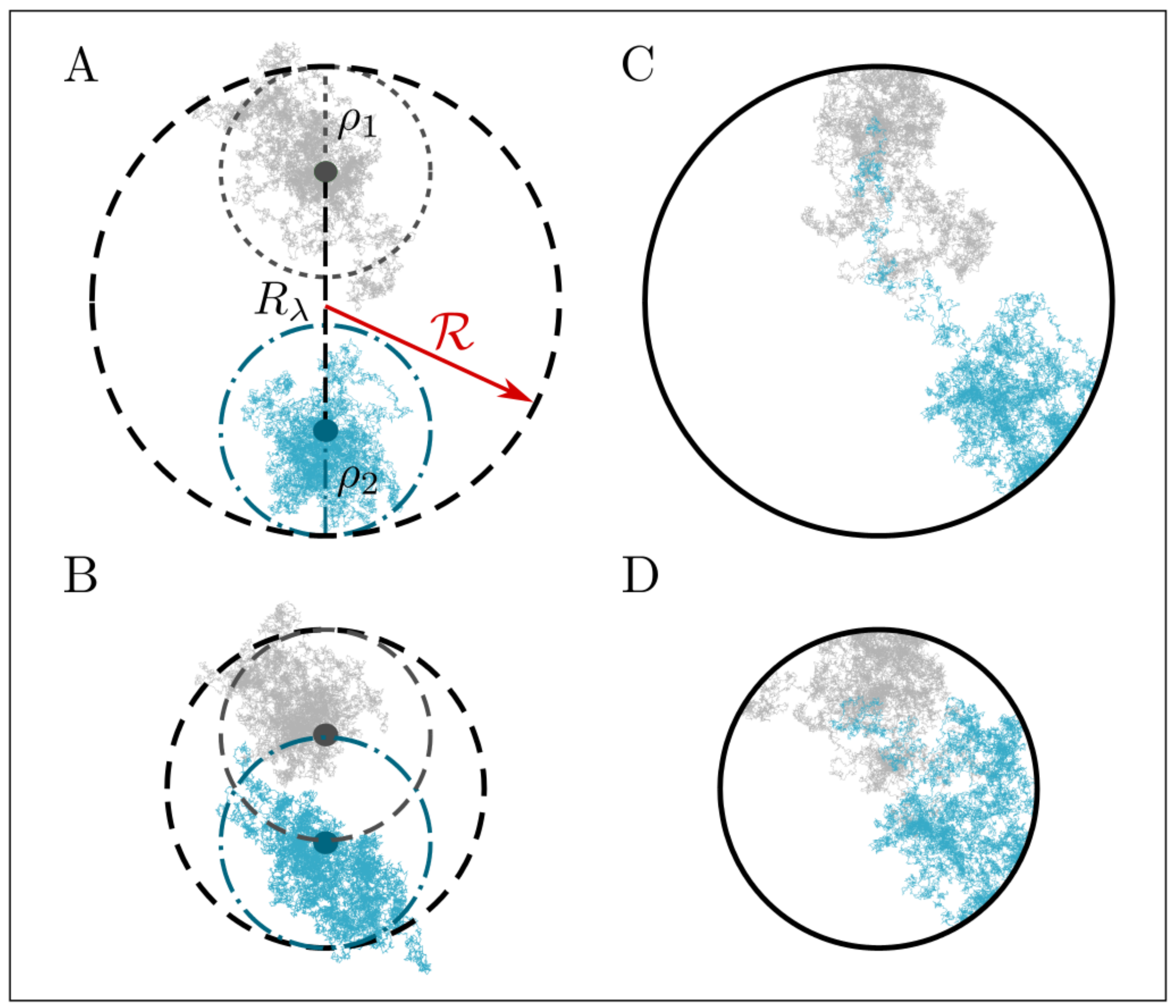}
        \caption{Sample trajectories for OU (A, B) and RBM (C, D). For the OU panels, the gray dashed and blue dashed-point circles delimit individual home ranges. The colored circles mark the home-range centers and the black dashed circle limits the population range, obtained from Eq.~(\ref{bm-joint}). Movement parameters are identical for both OU panels (A and B) other than the spatial distribution of the home ranges through $R_\lambda$. In the RBM panels, the encounter arena is limited by the solid circle, representing the reflecting boundary conditions.}\label{fig:trajectories}
\end{figure}

\textit{Individual home range and population range: definitions.} We define the home-range area as the smallest region of the space in which the probability of finding an OU individual is equal to an arbitrarily chosen quantile, $h$. Due to movement isotropy, individual home-ranges that result from this definition are circles with a radius that can be obtained by integrating the individual position PDF,
\begin{equation}\label{eq:defhomerange}
 \frac{1}{2\pi \sigma^2} \int_0^\rho \int_0^{2\pi} rdrd\theta {\rm e}^{-r^2/2\sigma^2} = h,
\end{equation}
 and therefore
\begin{equation}\label{eq:rho}
 \rho = \sqrt{2\sigma^2 \ln\left(\frac{1}{1-h}\right)} = \sqrt{g\tau}K,
\end{equation}
where $K\equiv\sqrt{-\ln(1-h)}$. For all the results presented here, we fix $h=0.95$, which is the conventional value used in the home range literature. Our results, however, are not qualitatively sensitive to changes in $h$. Importantly, even though the movement is restricted to an area, the individual position PDF has infinite support and therefore OU allows for occasional excursions in which animals leave their home range (Fig.~\ref{fig:trajectories}A, B), which is consistent with the widely-cited conceptual definition of home range introduced in \cite{Burt1943}. In contrast, when individuals are constrained within reflecting boundaries, such excursions are not possible (Fig.~\ref{fig:trajectories}C, D).

\begin{figure}
    \centering
        \includegraphics[width=0.55\textwidth]{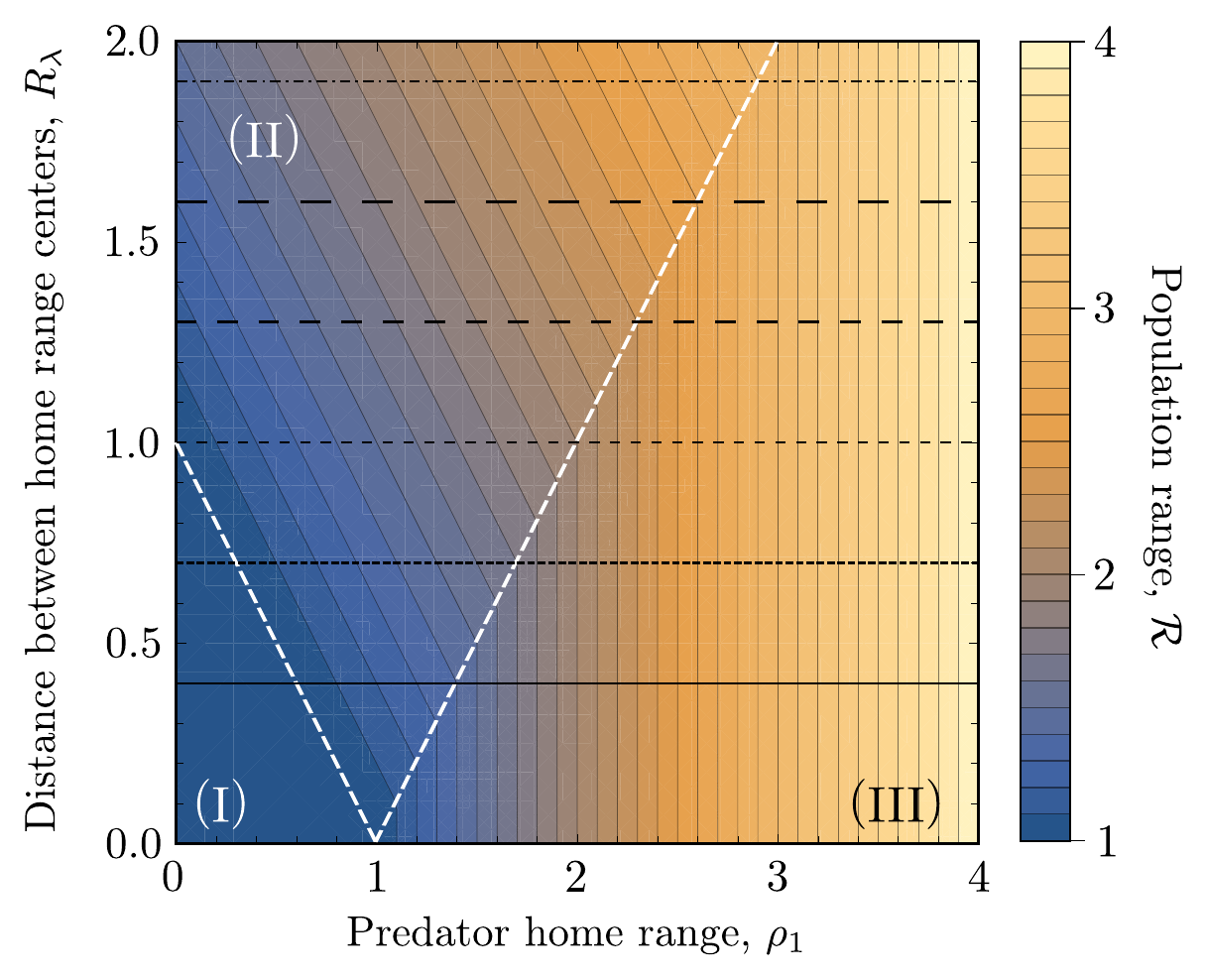}
        \caption{Contour plot of the population range as a function of the distance between home range centers, $R_\lambda$ and predator home range radius, $\rho_1$. The prey home range radius is kept constant, $\rho_2=1$. The white dashed lines mark the transitions in the branch of Eq.~(\ref{radius}) that defines the radius of the population range, $\mathcal{R}$. The horizontal black lines mark the values of $R_\lambda$ used in Fig.~\ref{fig:aier-bmou}.  }
        \label{fig:poprange}
\end{figure}

Based on this definition of home range, we define the population range as the circular region whose radius, $\mathcal{R}$, is equal to the radius of the smallest circular area that contains the stationary home range of both the OU prey and the OU predator (Fig. \ref{fig:trajectories}A, B). Depending on the distance between home range centers and the movement parameters, the population range radius is,
\begin{equation}\label{bm-joint}
 \mathcal{R}= \mbox{max}\left(\rho_1, \rho_2, \frac{(\rho_1 + \rho_2 + R_\lambda)}{2} \right).
\end{equation}
and it thus differs from individual home ranges except for the very particular case in which $\rho_1=\rho_2$ and $R_\lambda = 0$. With some algebra, we can obtain the conditions for which  the population range is defined either by the prey home range (Region I), by the predator home range (Region III), or by a combination of both and the distance between home range centers (Region II) (Fig.~\ref{fig:poprange}),
\begin{equation}\label{radius}
\mathcal{R} =\left\{\begin{array}{cll}
 \rho_2 & \mbox{if} \ \ \rho_1 < \rho_2 - R_\lambda  & \mbox{Region (I)} \\
 \frac{\rho_1 + \rho_2 + R_\lambda}{2} & \mbox{if} \ \ \rho_2 - R_\lambda < \rho_1  <  \rho_2 + R_\lambda & \mbox{Region (II)}\\
 \rho_1 & \mbox{if} \ \ \rho_1 >  \rho_2 + R_\lambda & \mbox{Region (III)}\\
\end{array}
\right.
\end{equation}

\subsubsection*{Reflected Brownian Motion}\label{sec:BBM}
RBM individuals move according to a pure BM enclosed within a finite container with reflecting boundaries that defines the population range. Because RBM does not account for home range affinity, the prey and the predator explore the whole population range in which they move and both individual home ranges are equal to each other and to the population range. In contrast to OU models, due to the reflecting boundary conditions, the PDF for individual RBM positions is not Normal in the long-term. In general, it is a function of the geometry of the population range \citep{Harrison1985}, but for the circular population ranges considered here, the stationary PDF for RBM positions is a uniform distribution defined inside the population range
\begin{equation}\label{rbm-pos}
 f_{\mbox{\tiny{RBM}}}(x,y,t\rightarrow\infty) =\left\{\begin{array}{cl}                                                 
\frac{1}{\pi \mathcal{R}^2} & \mbox{if} \ \ \sqrt{x^2+y^2} \leq \mathcal{R} \\
 0 & \mbox{otherwise}                                                                                                     \end{array}
\right.
\end{equation}

Therefore, RBM individuals use space uniformly for circular and other sufficiently smooth boundary geometries, while more complicated home range geometries might lead to nonuniform space use. It is, however, important to note that, in those cases, the more often visited regions would be determined by the shape of the reflecting borders rather than by intrinsic properties of individual movement or the landscape of resources in which individuals move \citep{Harrison1985}.

\section*{Results}

\subsection*{Pairwise distance distributions and mean encounter rate} \label{sec:genresults}

\subsubsection*{Ornstein-Uhlenbeck}
Because the position of each individual, $\bm{z}$, is normally distributed, the difference between them, $\Delta\bm{z}=(\Delta x, \Delta y)$, follows a Normal distribution as well. For this new variable, its mean is equal to the difference between the mean position of each individual, and the variance is equal to the sum of the variances in the position of each individual. In addition, due to movement isotropy, each component of $\Delta \bm{z}$ has the same variance, $\sigma_1^{2}+\sigma_2^{2}$, and we can define a nondimensional squared-distance $u(t)\equiv r^2(t)/\sigma_{r}^2$, where $\sigma_{r}^2\equiv\sigma_1^{2}+\sigma_2^{2}$ and $r^2(t)=\Delta x^2(t) + \Delta y^2(t)$. $u$ is a noncentral chi-squared variable with nondimensional noncentrality parameter $\tilde{\Lambda}\equiv\Lambda/\sigma_{r}^2 = (\mu_{\Delta x}^{2}+\mu_{\Delta y}^{2})/\sigma_{r}^2$, where $\mu_{\Delta x}$ and $\mu_{\Delta y}$ are the mean differences in the position of the individuals in the $x$ and $y$ coordinate respectively. They are calculated as $\mu_{\Delta x}=\mu_{x,1} - \mu_{x,2}$ and $\mu_{\Delta y} = \mu_{y,1} - \mu_{y,2}$ respectively. Because the nondimensional squared-distance, $u$, is a noncentral chi-squared variable, its PDF is
\begin{equation}\label{pdf-2Ddist}
 f_{\mbox{\tiny{OU}}} (u(t),\tilde{\Lambda}(t))= \frac{1}{2}\exp\left(\frac{-(u+\tilde{\Lambda})}{2}\right) I_0\left(\sqrt{\tilde{\Lambda} u}\right),
\end{equation}
where we have suppressed the time dependence in $u$ and $\tilde{\Lambda}$ on the right side of the equation for simplicity in the notation. $I_0$ is the modified Bessel function of the first kind and order $0$. The shape of $f(u;\tilde{\Lambda})$ for different numerical values of both $\tilde{\Lambda}$ and $\sigma_{r}^2$ is shown in Fig.~\ref{fig:distpdf}.

Inserting Eq.~(\ref{pdf-2Ddist}) in the definition of the mean encounter rate and solving the integral in Eq.~(\ref{eq:mean-encounter}) (see Appendix \ref{app-der} for detailed calculations), we obtain the mean encounter rate
\begin{equation}\label{eq:encOU}
  \tilde{\mathcal{E}}_{\mbox{\tiny{OU}}} (t,q)= \frac{\gamma}{\pi q^2}\left[ 1 - Q_1\left(\frac{\sqrt{\Lambda}(t)}{\sigma_r(t)},\frac{q}{\sigma_r(t)}\right)\right],
\end{equation}
where $Q_1$ is the Marcum-$Q$-function \citep{bocus2013approximation}. Importantly, because we have not made any assumption about the particular form of $\Lambda$ and $\sigma_r$, Eq.~(\ref{eq:encOU}) is a general expression that gives the mean encounter rate for any isotropic movement model in which the position of the individuals have Gaussian PDF with covariances equal to zero. Although Eq.~(\ref{eq:encOU}) provides an exact expression for the mean OU encounter rate, its dependence on the Marcum-Q-function makes it difficult to develop a quantitative understanding of how encounter rates are determined by the movement and perception parameters. To eliminate the Marcum-Q-function from the encounter rate, we study Eq.~(\ref{eq:encOU}) in the short perception limit $(q\ll\mathcal{R})$. Following the calculations provided in Appendix \ref{approximation-OU}, we obtain
\begin{equation}\label{eq:app-encOU}
   \tilde{\mathcal{E}}_{\mbox{\tiny{OU}}}(t,q) \approx \frac{\gamma \left[8\sigma_r^4(t) + q^2(\Lambda(t) - 2\sigma_r^2(t)) \right]}{16\pi\sigma_r^6(t)} \exp\left(-\frac{\Lambda(t)}{2\sigma_r^2(t)}\right).
\end{equation}

Finally, in the local-perception limit $(q=0)$, the mean encounter rate can be written in terms of the overlap between each individual's home range, measured either through the Bhattacharyya coefficient (BC) \citep{Fieberg2005,winner2018statistical} or through the scalar product of both individual-position PDFs, $f_1\!\cdot\! f_2$. $f_1$ and $f_2$ are, respectively, the PDFs for the position of the predator and the prey, and the scalar product, denoted by the symbol $\cdot$, is defined as the spatial integral of the product of both PDFs. 

In terms of the BC, the OU mean encounter rate is,
\begin{equation}\label{eq:OUenBC}
\tilde{\mathcal{E}}_{\mbox{\tiny{OU}}}(t;q=0) = \frac{\gamma \sigma_r^2(t)}{8\pi \sigma^2_1(t) \sigma^2_2(t)} BC^2(t),
\end{equation}
where the BC is a function of the movement parameters of the prey and the predator. Therefore, after a transient period in which the home range overlap and hence the BC changes with time, the BC reaches a stationary, constant value. In terms of the scalar product of the individual-position PDFs we get
\begin{equation}\label{eq:OUenSP}
\tilde{\mathcal{E}}_{\mbox{\tiny{OU}}}(t;q=0) = \gamma \ f_1\!\cdot\!f_2.
\end{equation}

\subsubsection*{Reflected Brownian Motion}
For RBM, individuals perform BM within a reflecting population range of radius $\mathcal{R}$. Therefore, to obtain the position PDF for each individual at any time, we need to solve the diffusion equation on a disk of radius $\mathcal{R}$ with reflecting boundary conditions, which is mathematically more challenging than the OU case. At short times, however, the effect of the reflecting boundaries is negligible and RBM converges to BM. Therefore, the pairwise distance distribution is given by Eq.~(\ref{pdf-2Ddist}) with noncentrality parameter equal to the distance between the initial position of each individual and rescaling variance $\sigma_{r}^2= (g_1+g_2) t$ (see Appendix \ref{app-BM}). In the stationary limit, because the use of space is uniform and population ranges are circular, the pairwise distance PDF is equal to the PDF for the distance between two random points within a disk of radius $\mathcal{R}$  \citep{Garcia-Pelayo2005},
\begin{equation}\label{bm-dist}
 f_{\mbox{\tiny{RBM}}}(r,t\rightarrow\infty) =\left\{\begin{array}{cl}                                                 
\frac{4r}{\pi \mathcal{R}^2}{\rm ArcCos}\left(\frac{r}{2\mathcal{R}}\right) - \frac{2r^2}{\pi \mathcal{R}^4}\sqrt{\mathcal{R}^2 - \frac{r^2}{4}} & \mbox{if} \ \ r \leq 2\mathcal{R}, \\
 0 & \mbox{otherwise},                                                                                                  \end{array}
\right.
\end{equation}
which is shown in Fig. \ref{fig:distpdfbm} for various values of $\mathcal{R}$. Inserting the PDF from Eq.~(\ref{bm-dist}) in the definition of the mean encounter rate, Eq.~(\ref{eq:mean-encounter}), we obtain the stationary RBM mean encounter rate, $\tilde{\mathcal{E}}_{\mbox{\tiny{RBM}}}$,
\begin{align}\label{eq:meanenc-bbm}
 \tilde{\mathcal{E}}_{\mbox{\tiny{RBM}}}&(t\rightarrow\infty, q) = \nonumber \\
 =& \frac{\gamma}{4\pi^2 q^2 \mathcal{R}^3} \left[ 8\mathcal{R}\left(R^2{\rm ArcCsc}\left(\frac{2\mathcal{R}}{q}\right) 
+ q^2 {\rm ArcSec}\left(\frac{2\mathcal{R}}{q}\right) \right) - q^3 \sqrt{4-\frac{q^2}{\mathcal{R}^2}} + 2q\mathcal{R}\sqrt{4\mathcal{R}^2-q^2} \right].
\end{align}

Finally, we can also obtain the approximate encounter rate in the short-perception limit (see Appendix \ref{approximation-BM} for detailed calculations),
\begin{equation}\label{bm-enc-app-main}
 \tilde{\mathcal{E}}_{\mbox{\tiny{RBM}}}(t\rightarrow\infty, q) \approx \frac{\gamma}{\pi\mathcal{R}^2}\left( 1 - \frac{4 q}{3\pi \mathcal{R}}\right).
\end{equation}

In summary, OU encounter rates depend on four spatial scales, to which we can give different ecological meanings. First, the predator perceptual range, $q$, contains all the information about the encounter process itself, assuming that the encounter kernel is fixed. Second, the distance between home range centers, $R_\lambda$ provides information about the spatial distribution of home ranges. Finally, each of the home range radii, $\rho_1$ and $\rho_2$, contains information about individual movement process and habitat use. RBM encounter rates, in contrast, only depend on two spatial scales: the predator perceptual range, $q$ and the population range, $\mathcal{R}$. This is a very important difference between OU and RBM encounters: in RBM, the long-term encounter rate is completely determined by the population range (if the perceptual range is constant), but in OU the long-term encounter rate depends on the combination of values for $\rho_1$, $\rho_2$ and $R_\lambda$ that leads to that particular population range.

\subsection*{Short-term encounter rates} 

\subsubsection*{Ornstein-Uhlenbeck} \label{sec:transient-ou}

In this section, we investigate the transient behavior of the encounter rate, that is, before the OU moments reach their stationary values. We consider a pair of individuals that follow identical OU models ($g_1=g_2\equiv g$ and $\tau_1=\tau_2\equiv \tau$) departing from their individual-specific home range centers. Mathematically, this initial condition is given by ${\bm \mu} (0)=(\lambda_x, \lambda_y)$ and $\sigma^2(0)=0$. Therefore, $R_\lambda$ is both the mean and the initial distance between individuals. Inserting these initial conditions in Eqs.~(\ref{mean-pos}) and (\ref{var-pos}), the mean position of each individual is constant and fixed at its home-range center whereas the variance increases monotonically from zero until it reaches its steady value. The noncentrality parameter and the rescaling variance are,
\begin{eqnarray}
\Lambda &=& R_\lambda^2 \label{lambdapar} \\
\sigma_r^2(t) &=& g\tau \left(1- {\rm e}^{-2t/\tau} \right). \label{sigmapar}
\end{eqnarray}

\begin{figure}
    \centering
        \includegraphics[width=\textwidth]{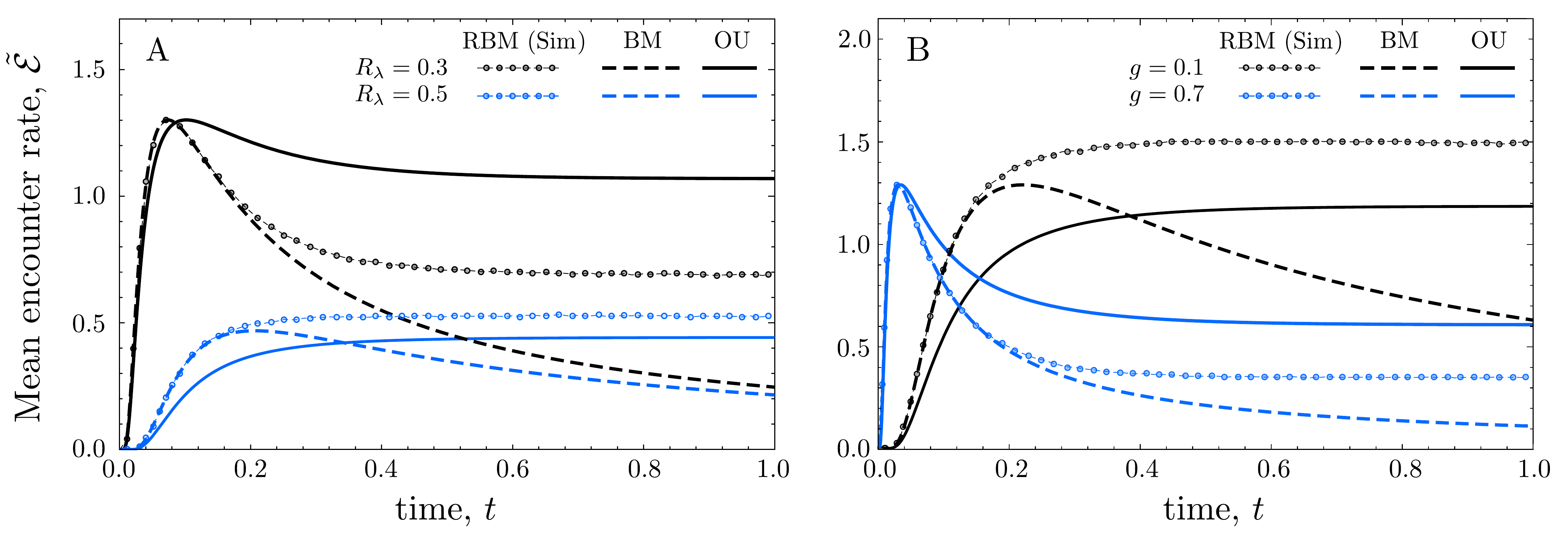}
        \caption{Transient dynamics in the mean encounter rate for OU (analytical solution, solid lines), BM (analytical solution, dashed lines), and RBM (numerical averages over $5\times10^7$ realizations with identical initial conditions, dotted lines). In each panel, curves with the same color are obtained starting from the same initial conditions and using the same movement parameters. In panel A, the intensity of the stochasticity in individual movement is kept constant, $g=0.3$, and blue and black curves differ in the distance between home range centers, $R_\lambda$; in panel B, contrarily, the position of the individual home ranges is constant, $R_\lambda=0.3$, and movement stochasticity, $g$, varies. Other parameters remain constant and take the same value in both panels, $\tau=0.3$ (for OU models) and $q=0.05$.}
        \label{fig:transient}
\end{figure}

Inserting Eq.~(\ref{lambdapar}) and (\ref{sigmapar}) into Eq.~(\ref{eq:encOU}), we obtain a closed expression for the mean encounter rate that only depends on predator perceptual range, $q$, and the various movement parameters. Because we start with deterministic initial conditions, the initial encounter rate is zero unless both individuals have the same home range center. As time elapses, however, the position of each individual becomes uncertain due to movement stochasticity, and the probability that individual trajectories cross with each other increases. Consequently, the mean encounter rate increases too. The mean encounter rate grows monotonically until it stops at a constant, stationary value if the overlap between home ranges is low ($R_\lambda$ large, $g$ small). The mean encounter rate shows, however, a maximum at an intermediate time during the transient regime when the overlap between home ranges is larger ($R_\lambda$ small, $g$ large)  (Fig.~\ref{fig:transient}). Moreover, if the distance between home-range centers is kept constant, the intensity of the stochastic movement component, $g$, controls a tradeoff between short-term and long-term mean encounter rate. More stochastic movements favor early encounters because individuals spread faster from their initial positions and are more likely to encounter each other quickly. However, in the long term, because home ranges are larger, the encounter probability is smaller (Fig.~\ref{fig:transient}B).

\subsubsection*{Reflected Brownian Motion}

Next, we compare OU mean encounter rates with those obtained for a pair of RBM individuals that move within the population range, $\mathcal{R}$, defined by the OU movement parameters with which we want to make the comparison. Therefore, we calculate the population range of each pair of OU individuals considered in the previous section and use that value to constrain the movement of a pair of RBM individuals. Because an analytical expression for the time-dependent pairwise distance PDF is not accessible for RBM, we perform numerical simulations of the encounter process. We find that while some aspects of the qualitative behavior are similar to OU, others differ in important ways. For fixed movement parameters, when the population range is large and individuals start far from each other, the encounter rate is a monotonically increasing function of time. Conversely, if the population range is small and individuals start close to each other, the maximum overlap between trajectories, and hence the maximum encounter rate, is reached at an intermediate time (Fig.~\ref{fig:transient}A). However, the transition from monotonically increasing encounter rates to first increasing and then decreasing encounter rates does not necessarily occur at the same $R_\lambda$ for OU and RBM models. For a fixed initial distance, $R_\lambda$, the stochasticity in the movement also controls a tradeoff between short-term and long-term encounter rate qualitatively similar to the one featured by OU encounters (Fig.~\ref{fig:transient}B). Finally, for completeness, we also study the pure BM case, in which individual movement is not constrained by the population range (see Appendix \ref{app-BM} for details). In the short-time limit RBM and BM provide identical results because the effect of the boundary conditions on RBM is negligible and the movement statistics in both cases are, consequently, the same. As border reflections become more important in RBM, however, BM and RBM encounter rates diverge from each other: $\tilde{\mathcal{E}}_{\mbox{\tiny{RBM}}}$ stabilizes to its stationary value whereas $\tilde{\mathcal{E}}_{\mbox{\tiny{BM}}}$ decays to zero.

\subsection*{Steady state encounter rates}

Next, we evaluate the encounter rate in the stationary limit, when the moments of the movement models have reached a constant value.  We first analyze the dependence of OU encounter rates on predator home range size and perceptual range and then investigate whether the main features of range residency, nonuniform use of space and restricted use of the population range, lead to qualitative differences between OU and RBM-based encounter rates.

\subsubsection*{Ornstein-Uhlenbeck encounter rate}\label{sec:ou-steady}
In this stationary regime ($t\rightarrow\infty$), the rescaling variance and the noncentrality parameter of the nondimensional squared-distance PDF are constant
\begin{eqnarray}
\Lambda &=& R_\lambda^2, \label{lambdapar-st} \\
\sigma_r^2 &=& \frac{\rho_1^2 + \rho_2^2}{2 K^2}, \label{sigmapar-st}
\end{eqnarray}
where we have used the definition of the home range radius, $\rho = \sqrt{g\tau}K$, so the pairwise distance PDF depends explicitly on all the spatial scales that, together with the perceptual range, determine the encounter rate.

\begin{figure}
    \centering
        \includegraphics[width=\textwidth]{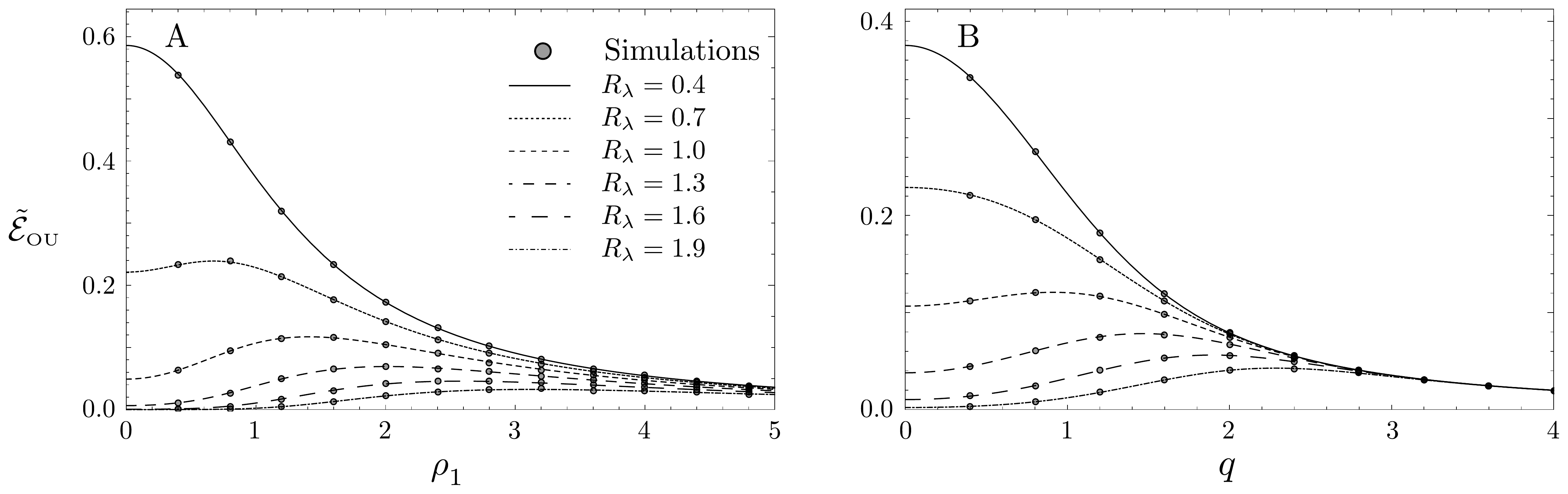}
        \caption{OU mean encounter rate versus $\rho_1$ with $q=0.1$ (A), and versus $q$ with $\rho_1 = 1$ (B). Symbols correspond to numerical simulations and $\rho_2 = 1$ in both panels.}
        \label{fig:aier-bmou}
\end{figure}

For a first analysis of the encounter statistics, we keep the home range radius of the prey constant, $\rho_2 = 1$, and study the behavior of the encounter rate for different distances between home range centers, $R_\lambda$, perceptual ranges, $q$, and predator home range radii, $\rho_1$. The distance between home range centers, $R_\lambda$, quantifies the habitat configuration; the perceptual range, $q$, contains all the information about the encounter process for a given shape of the encounter kernel; and $\rho_1$ informs us about predator space use. In addition, because $\rho_2=1$, $\rho_1$ gives the size of the predator home range relative to the prey's. Finally, each set of values for $\rho_1$, $\rho_2$, and $R_\lambda$ define a population range, $\mathcal{R}$, according to Eq.~(\ref{bm-joint}) (Fig.~\ref{fig:poprange}). Varying either $R_\lambda$ or $\rho_1$ changes the home range overlap and hence the encounter rate. Increasing $R_\lambda$ with constant predator home range radius, $\rho_1$, and perceptual range, $q$, decreases the overlap between home ranges and, as a result, decreases the encounter rate (Fig.~\ref{fig:aier-bmou}A, B). 

The predator home-range radius has different effects on home-range overlap depending on the value of $R_\lambda$. If home range-centers are far from each other (large $R_\lambda$), as predator home range increases with respect to prey home range (increasing $\rho_1$) the overlap between home ranges and thus the encounter rate increases. However, if the predator home range continues to grow and becomes much larger than the prey's ($\rho_1 \gg 1$), encounters start becoming rarer and encounter rates decrease (Fig.~\ref{fig:aier-bmou}A except solid line). When the distance between home-range centers is short (small $R_\lambda$), larger predator home ranges immediately make encounters rarer and thus the mean encounter rate is a monotonically decreasing function of $\rho_1$ (solid line in Fig.~\ref{fig:aier-bmou}A). We therefore observe a transition in the predator home-range size that maximizes the encounter rate, that is, on the amount of territory that predator should explore to maximize its predation success. If predator and prey home ranges are close to each other or the prey home range is sufficiently large, then the optimal predator home range is $\rho_1 =0$, which represents an ambush predation strategy. On the contrary, if the prey's home range is small or located far away from the home range center of the predator, then the encounter rate is maximum for some predator mobility (see Appendix \ref{app:eff-homerange} for detailed calculations). This richness of behaviors in the encounter rate is due to the home ranging features introduced by OU movement. Using Eq.~(\ref{bm-joint}), any pair of values  $(\rho_1$, $R_\lambda$), together with the constant $\rho_2=1$, define a population range $\mathcal{R}$ that contains the home ranges of both individuals (see Fig.~\ref{fig:poprange}). If we neglect the effect of range residency and calculate the encounter rate between a pair of RBM individuals moving within that same population range, we find that the RBM encounter rate is a decreasing function of $R_\lambda$ and $\rho_1$ (see Appendix \ref{app:rbm-steady} for more details). This occurs because larger values of $R_\lambda$ and $\rho_1$ result in larger population ranges (see transects in Fig.~\ref{fig:poprange}), and the RBM encounter rate is a monotonically decreasing function of $\mathcal{R}$.

If $\rho_1$ remains constant and the perceptual range $q$ varies, the OU encounter rates behave qualitatively similar to the fixed $q$ and variable $\rho_1$ case (Fig.~\ref{fig:aier-bmou}B). However, in the limit $q\rightarrow\infty$, the pairwise CDF reaches $1$ and all the encounter rates decay as $\sim 1/(\pi q^2)$ regardless of the distance between home range centers and regardless of the movement model. Therefore, for short perceptual ranges, the OU encounter rate may be either a decreasing or an increasing function of $q$. In the former case, local perception $(q=0)$ leads to a maximum of the encounter rate. In the latter, $q=0$ is a local minimum and the encounter rate might be maximal for intermediate perceptual ranges (See Appendix \ref{app:effq} for details). Finally, because $\rho_1$, $\rho_2$, and $R_\lambda$ are kept constant for each curve in Fig..~\ref{fig:aier-bmou}B, the population range remains constant as well. Hence, if we neglect the effect of range residency and enclose a pair of RBM individuals within the population range defined by the set of values for $(\rho_1, \rho_2, R_\lambda)$ used in each curve, the encounter rate loses its complex dependence on the perceptual range and it becomes a monotonically decreasing function of $q$ (see Appendix \ref{app:rbm-steady} for more details).

\subsubsection*{OU-vs-RBM: the role of nonuniform space use}

\begin{figure}[!ht]
    \centering
        \includegraphics[width=0.99\textwidth]{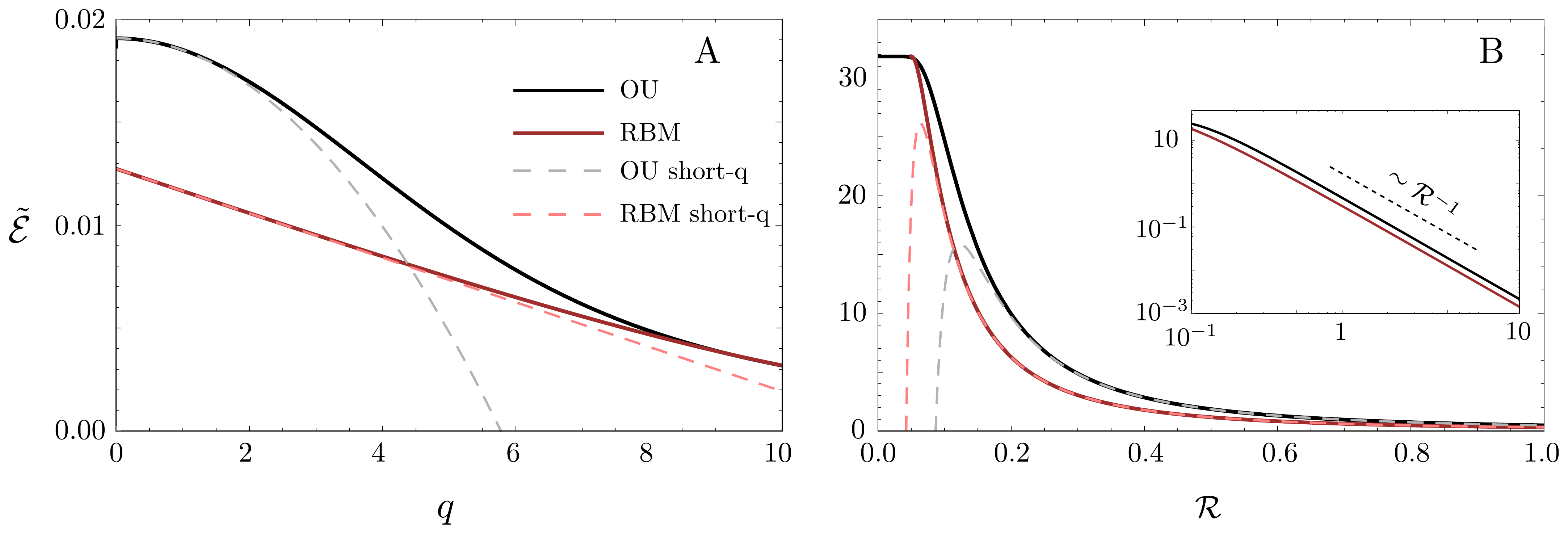}
        \caption{OU and RBM mean encounter rates as a function of the perceptual range, $q$, with population range $\mathcal{R}=5$ (A); and as a function of the population range with $q=0.1$ (B). Solid lines correspond to the exact mean encounter rate and dashed lines to the short-$q$ approximation. The inset of panel B shows the asymptotic decay of the OU and RBM mean encounter rate (log-log plot). In every panel: $R_\lambda = 0$, and  $\mathcal{R}\equiv\rho_1=\rho_2$ so that both OU and RBM explore the whole population range.}\label{fig:nonuniformspaceuse}
\end{figure}

To isolate the effect of nonuniform space use on encounter rates from other factors, such as individual home ranges being smaller than the population home range, we first consider the case in which the home range of both the prey and predator are equal to each other and to the population range, $\mathcal{R}$. This only occurs if the prey and predator individual position PDFs are identical to each other (i.e., they have the same mean and variance), which leads to $\Lambda = 0$ and $\sigma^2_r = \rho^2/K^2$. Notice that we have omitted the individual subscript in the notation of the home range radii and used $\rho\equiv\rho_1=\rho_2$. Substituting these values for the noncentrality parameter and the rescaling variance in Eq.~(\ref{eq:app-encOU}), we obtain, in the short-perception limit,
 \begin{equation}
    \tilde{\mathcal{E}}_{\mbox{\tiny{OU}}}(t\rightarrow\infty,q) \approx \frac{\gamma K^2}{2\pi \mathcal{R}^2}\left(1-\frac{K^2 q^2}{4\mathcal{R}^2}\right),
 \end{equation}
 where we have already used that $\rho \equiv \mathcal{R}$. Therefore, if both individuals occupy the whole population range but space use is nonuniform, the encounter rate decreases with the square of the perceptual range (Fig.~\ref{fig:nonuniformspaceuse}A). This behavior is different than the linear decay of Eq.~(\ref{bm-enc-app-main}) for RBM encounters, in which space use is uniform. Therefore, even though individual home ranges are identical to each other and to the population range for the OU and the RBM setups considered in this section, the nonuniform use of space featured by OU movement leads to a faster decay of the encounter rate as perceptual range increases. Moreover, because OU individuals have affinity to their home range centers and both prey and predator home range centers are located at the same position, the OU encounter rate is always larger than the RBM encounter rate (Fig.~\ref{fig:nonuniformspaceuse}A).
 
Regarding the decay of the encounter rate with the population range  (Fig.~\ref{fig:nonuniformspaceuse}B), we first observe that OU and RBM take the same values for small population ranges, because the home ranges are small and the effect of home range affinity in the encounter rate is negligible. As the population range becomes larger, however, the RBM encounter rate decays faster than the OU encounter rate. The lack of home-range affinity in the RBM case allows individuals to explore the borders of the population range more frequently and thus to be farther from each other. For very large population ranges, encounter rates from the two models tend to converge together because the intensity of OU home-range affinity is weaker.  However, they never take the same value, because both OU and RBM decay asymptotically as $\mathcal{R}^{-1}$ when $\mathcal{R} \gg q$ (inset in Fig.~\ref{fig:nonuniformspaceuse}B).

\subsubsection*{OU-vs-RBM: the role of restricted use of the population range}

\begin{figure}[!ht]
    \centering
        \includegraphics[width=\textwidth]{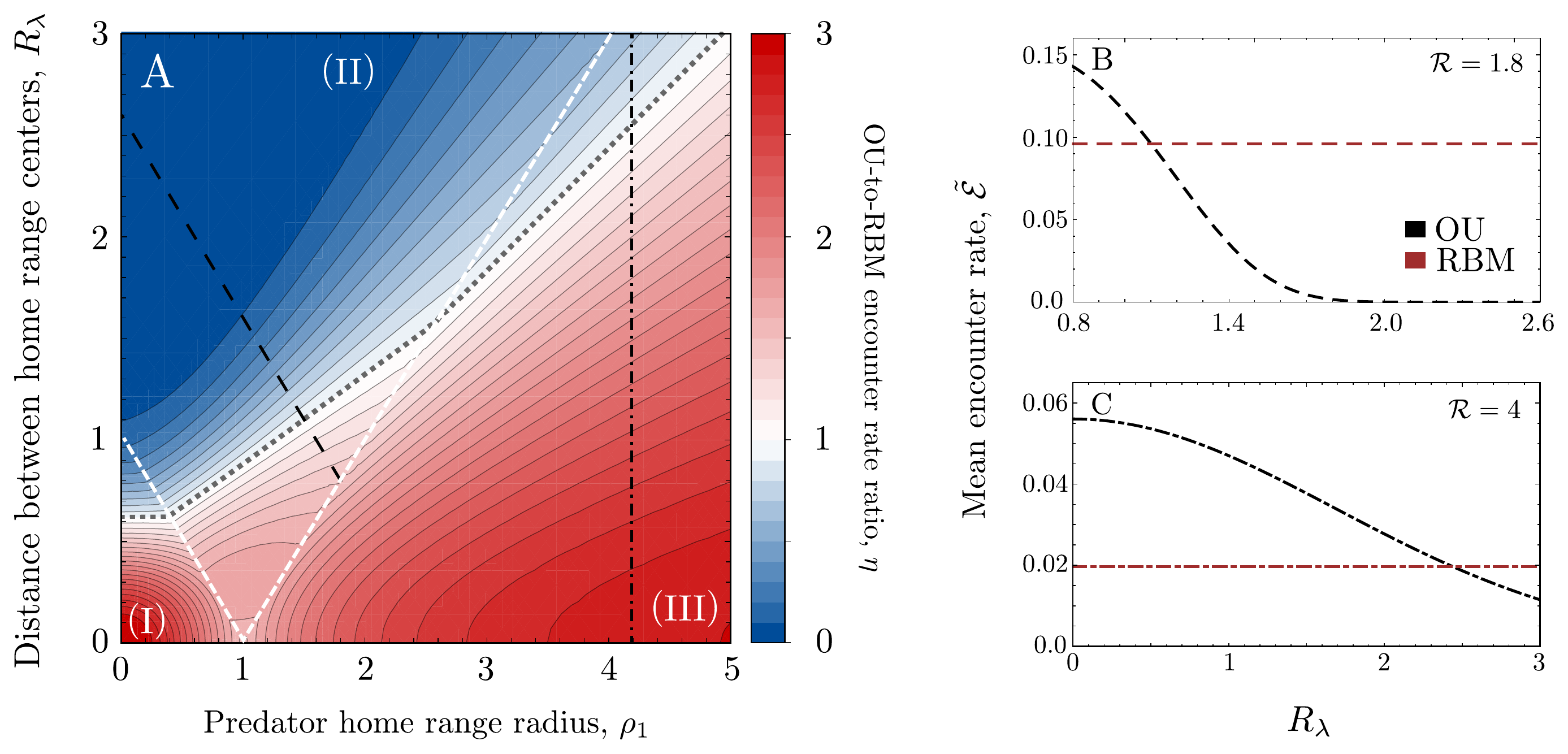}
        \caption{A) OU-to-RBM mean encounter-rate ratio for varying distance between OU home-range centers, $R_\lambda$, and OU predator home range radius, $\rho_1$. The contour lines have a spacing equal to $0.1$; the thicker-dotted line represents the contour equal to $1$. The white-dashed lines mark the transitions in the branch of Eq.~(\ref{radius}) (I, II or III) that defines the radius of the population range, $\mathcal{R}$. The black lines (dashed and dot-dashed) mark transects of the parameter space with $\mathcal{R}$ constant. B, C) OU and RBM encounter rate versus $R_\lambda$ for each of the transects traced in panel A. The type of line in each of the transects of panel A is maintained to plot the encounter rates in these panels (panel B corresponds to region II and panel C to region III). The color code of panel B is maintained in C.}\label{fig:restrictedspaceuse}
\end{figure}

\begin{figure}[!ht]
    \centering
        \includegraphics[width=0.985\textwidth]{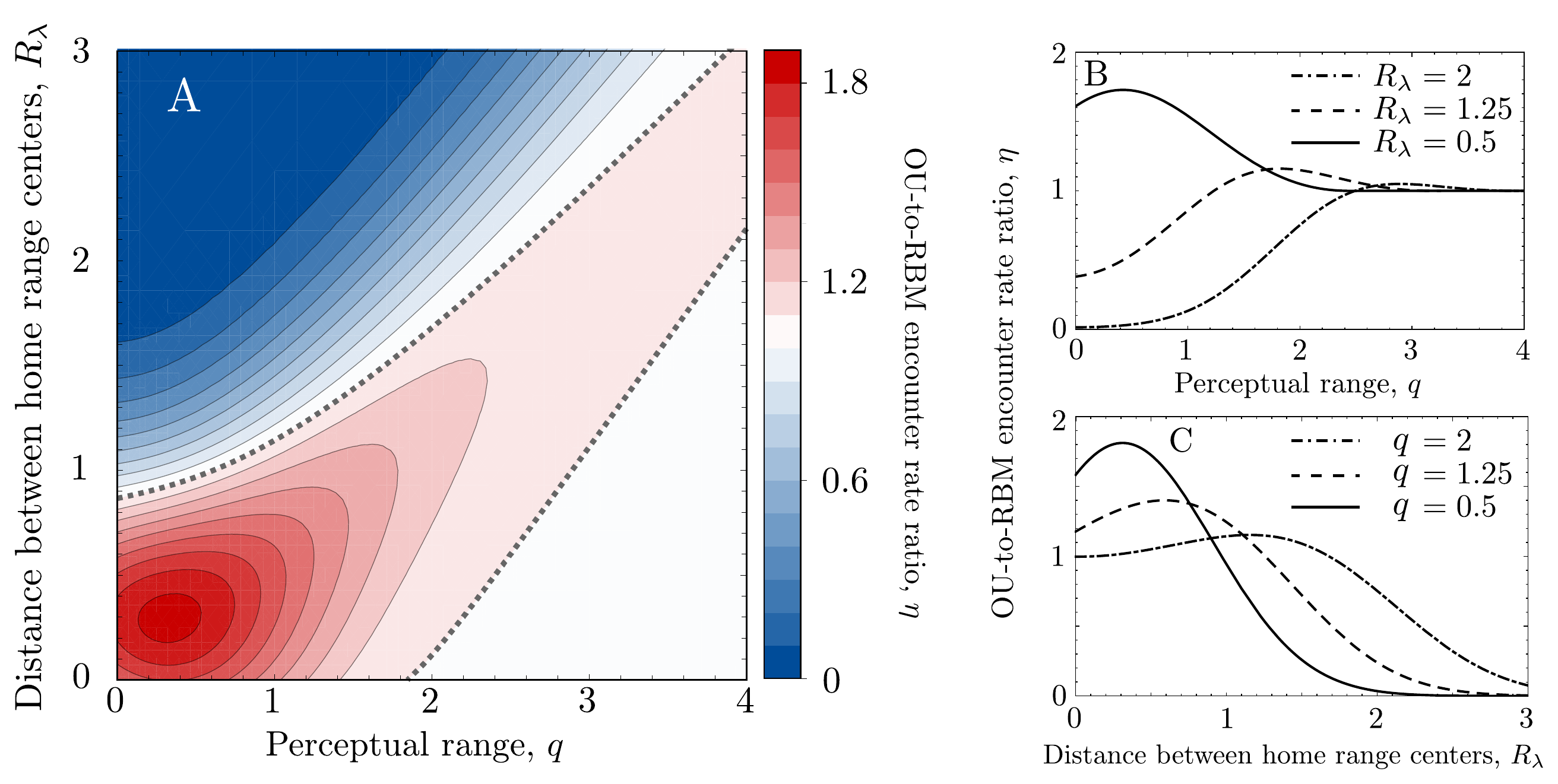}
        \caption{A) OU-to-RBM mean encounter rate ratio for varying distance between OU home-range centers, $R_\lambda$, and perceptual range, $q$. Both individual home-range radii are constant, $\rho_1=\rho_2=1$. The population range increases with increasing $R_\lambda$ but it always lies in region II as defined by Eq.~(\ref{hr-to-poprange}) (Fig.~\ref{fig:poprange}). The contour lines have a spacing equal to $0.1$ with the thicker-dotted line representing the contour equal to $1$. B) OU-to-RBM mean encounter rate ratio along three transects with constant $R_\lambda$. C) Same as panel B) but along transects with constant $q$.}\label{fig:restrictedspaceuse-q}
\end{figure}

The second feature of the OU movement model, as opposed to RBM, is that it allows individual home ranges to differ from the population range. This limits the area in which encounter may occur (i.e., from the entire population range to the overlap between the more restricted home ranges), which may introduce important qualitative differences between RBM and OU encounter statistics. In this section, we evaluate the ratio between the OU and the RBM mean encounter rate, $\eta\equiv\tilde{\mathcal{E}}_{\mbox{\tiny{OU}}}/\tilde{\mathcal{E}}_{\mbox{\tiny{RBM}}}$, over a range of conditions that include changing the distribution and sizes of home ranges within a population range at constant perceptual range, Fig.~\ref{fig:restrictedspaceuse}, and varying perceptual ranges on a constant spatial distribution of home ranges and home range sizes, Fig.~\ref{fig:restrictedspaceuse-q}. This analysis will therefore quantify the accuracy of approximating more realistic OU movement with RBM. More specifically, if $\tilde{\mathcal{E}}_{\mbox{\tiny{OU}}}/\tilde{\mathcal{E}}_{\mbox{\tiny{RBM}}}>1$, then RBM underestimates the effect of home ranging behavior on encounter rates; if $\tilde{\mathcal{E}}_{\mbox{\tiny{OU}}}/\tilde{\mathcal{E}}_{\mbox{\tiny{RBM}}}<1$, RBM overestimates the effect of home ranging behavior; finally, RBM is an accurate approximation to home ranging behavior if $\tilde{\mathcal{E}}_{\mbox{\tiny{OU}}}/\tilde{\mathcal{E}}_{\mbox{\tiny{RBM}}}\approx 1$.

First, we keep the prey home range, $\rho_2$, and the perceptual range, $q$, constant and allow the predator home range, $\rho_1$, and the distance between home ranges, $R_\lambda$, to change. As both $\rho_1$ and $R_\lambda$ change, the population range $\mathcal{R}$ also changes according to Eq.~(\ref{bm-joint}) (see Fig.~\ref{fig:poprange}). The way the population range changes with the distance between home ranges and the predator home range defines three regions in the $(R_\lambda, \rho_1)$ parameter space, delimited by the white-dashed lines in Fig.~\ref{fig:restrictedspaceuse}A. In region (I), the population range is equal to the prey home range; in region (II), the population range is a linear combination of both home-range radii and the distance between their centers; in region (III), the population range is equal to the predator home-range radius (see Fig.~\ref{fig:scheme} for schematic examples of how home ranges arrange within the population range in each case). To understand how restricted use of the population range impacts encounter rates, we need to move in each of these regions in a way that the population range remains constant despite changes in $R_\lambda$ and/or $\rho_1$. 

In region (I), the population range, $\mathcal{R}$, is equal to the prey home-range radius and thus remains constant in the entire region regardless of the value of $R_\lambda$ and $\rho_1$. Because the RBM mean encounter rate only depends on $\mathcal{R}$, it is also constant in the entire region. The OU mean encounter rate, however, depends on the size and the location of the predator home range. In general, when $\rho_1$ and $R_\lambda$ are small, range residency maintains individuals closer to each other and thus the mean encounter rate is larger for the OU model than for the RBM. As $R_\lambda$ and $\rho_1$ increase, the OU mean encounter rate decreases. Finally, for the largest $R_\lambda$ allowed in this region, the mean encounter rate is smaller for OU than for RBM because the predator home range is centered close to the boundary of the prey home range, which is rarely visited by the prey. In region (II), for the population range to remain constant, the size of the predator home range must decrease as the distance between home range centers increases (black-dashed line in Fig.~\ref{fig:restrictedspaceuse}A and Fig.~\ref{fig:poprange}). Along such a transect of constant population range, the OU-to-RBM mean encounter rate ratio decreases as $R_\lambda$ increases because the RBM encounter rate is constant and the OU encounter rate decreases monotonically as $R_\lambda$ increases and $\rho_1$ decreases simultaneously (Fig.~\ref{fig:restrictedspaceuse}B). In region (III), we observe the same trend in the OU-to-RBM mean encounter rate ratio as $R_\lambda$ increases for constant population range (Fig.~\ref{fig:restrictedspaceuse}C; notice that the population-range radius is equal to the predator home-range radius in this region). Finally, both in regions (II) and (III), the RBM encounter rate may transition from underestimating to overestimating OU encounter depending on the size of the population range.

Second, we keep the predator home-range radius constant and equal to $\rho_2$, $\rho_1=1$, and explore the effect of the perceptual range on the OU-to-RBM mean encounter rate ratio. Importantly, because individual home ranges are equal to each other, the population range is always defined by the branch (II) of Eq.~(\ref{radius}) and depends linearly on the distance between home-range centers,
\begin{equation}\label{hr-to-poprange}
\mathcal{R} = 1 + \frac{R_\lambda}{2}.
\end{equation}

The perceptual range is a central parameter for encounter rates and hence helps determine the accuracy of assuming simplified movement models. Perceptual ranges vary importantly across different species and within individuals of the same species  \citep{zollner1997landscape,zollner2000comparing,mech2002using,fletcher2013signal}. Therefore, for the same spatial distribution of home ranges, $R_\lambda$, and predator movement properties, $\rho_1$, a RBM approximation to home ranging might give accurate results for some species but inaccurate results for others. Our results show a more important disagreement between RBM and OU encounters (RBM either underestimates or overestimates the OU encounter rate) at short perceptual ranges regardless of the distance between home range centers. Shorter perceptual ranges require a more precise description of the encounter of the trajectories because the predator detects prey at very short distances. Therefore, neglecting range residency in movement models has a higher impact in this limit (Fig.~\ref{fig:restrictedspaceuse-q}). Finally, for large $q$, OU and RBM mean encounter rates are equal to each other because both decay as $\sim q^{-2}$ (see Fig.~\ref{fig:encasymptotic}).

%%%%%%%%%%%%%%%%%%%%%%%%%%%%%%%5
\section*{Discussion}

Understanding how different movement processes affect pairwise encounter rates is a key step in upscaling from individual movement behavior to population- and community-level consequences. Recent years have seen rapid developments in movement ecology, in particular with respect to statistical methods for extracting information from tracking data. However, encounter rate modeling has remained largely locked in the past, focusing on overly simplistic models such as Brownian motion that, while conceptually useful, do a poor job of describing real data. An accurate description of the pairwise encounters occurring within predator and prey populations is crucial for constructing well-grounded models of interacting populations. In such models, encounter terms have historically been based on the principle of mass action, which assumes that individuals follow RBM and encounter each other whenever their paths directly cross \citep{Hutchinson2007}. Because every individual is assumed equally likely to occupy any region of the space in RBM, predator-prey encounters are proportional to the product of the densities of the two species. More sophisticated versions of these models feature modified encounter terms that account for complexities including predator handling times or functional responses \citep{CSHolling1959,Oaten1975,Berryman1992}. However, all of these elaborations still rest on the principle of mass action and its underlying assumptions about individual movement. 

Here, we set the basis of an analytical theory for encounter rates between a pair of OU individuals, the simplest movement model that accounts for individual home ranges. Attraction to a home range center, as featured by OU, does not introduce important qualitative differences in the transient encounter rate as compared to RBM (Fig.~{fig:transient}). In both cases, the transient encounter rate grows monotonically when individuals start close to each other and the movement is dominated by deterministic attraction to the home range center, but the transient encounter rate has a maximum at intermediate times if the initial prey-predator distance is large and their trajectories more stochastic. Moreover, both for OU and RBM, stochasticity in individual trajectories controls a tradeoff between short-term and long-term encounter probability, with more stochastic movement patterns favoring earlier encounters. 

For steady-state encounter rates, uneven space utilization itself does not introduce qualitative differences between OU and RBM encounters either. However, neglecting uneven space utilization consistently leads to underestimations of the encounter probability, especially for short perceptual ranges and intermediate population ranges. In fact, RBM broadly fails to estimate encounter rates when perceptual ranges are short, providing either over or underestimated values depending on whether home ranges are close together or far apart, respectively. More importantly, fundamental qualitative differences between OU and RBM encounter models emerge from OU movement restricting individual ranges to a subregion of the population range. This sets a spatial scale, shorter than the population range, at which encounters can potentially occur. In RBM, in contrast, this scale is absent and individuals may interact all over the population range. Moreover, when coupled to a finite range of perception, this short scale at which OU encounters occur leads to complex functional shapes for the OU encounter rate (Fig.~\ref{fig:aier-bmou}). These manifest as intermediate optimal perceptual ranges and predator home-range sizes (see \cite{Martinez-Garcia2013, Martinez-Garcia2014, Martinez-Garcia2017, Fagan2017} for other scenarios where intermediate perceptual range turn out to be optimal). When both nonuniform space utilization and individually restricted ranges are turned off, as in RBM, this richness of behaviors is lost. 

Despite its unrealistic assumptions, one could pontentially view the RBM model as a pragmatic approximation to the more realistic OU model. Indeed, a simpler model that yields qualitatively correct behavior can sometimes be a useful tool for gaining insight \citep{Durrett2005}. However, we feel that this is not the case with RBM for three reasons. First, RBM is actually less analytically tractable than OU, as evidenced by our inability to obtain an exact, time-dependent expression for its mean encounter rate. Second, estimation procedures for RBM have not been developed, and likely will not be developed, because of the above-noted incongruence with realistic home-ranging behavior, combined with the rarity of situations where an individual's home range is completely defined by hard, reflecting boundaries \citep{Noonan2019}. So unlike OU, RBM cannot be rigorously applied to real tracking data. Finally, as we have demonstrated, the qualiative behavior of the RBM encounter model differs importantly from its OU-based counterpart in many cases. For these reasons, we unequivocally recommend OU as the best available framework for modeling encounter processes in the presence of home-ranging behavior.

By explicitly incorporating range residency in the underlying movement model, our analyses here have helped place encounter rate modeling on firmer empirical ground. However, the OU model itself still requires tracking data that are coarse enough to not show autocorrelated velocities, and many modern, high-resolution tracking datasets do not meet this criterion. The OUF process \citep{Fleming2014,Fleming2014b} is a generalization of OU that, in addition to range residency, also includes velocity autocorrelation. While OU does describe many real datasets well, the OUF model is, in our experience, a preferred model for range resident data. For example, in the comparative home range analysis performed by \citet{Noonan2019}, the OUF model was selected for 240 out of 369 datasets, while the OU model was selected for 128. Deriving encounter metrics for correlated velocity models like OUF will clearly be mathematically challenging, but this represents an important future opportunity building on our results.

Our main focus here has been quantifying encounter rates between a pair of predator and prey inviduals. Predator-prey interactions, however, are likely to occur among many individuals within populations of each species. For a single predator navigating a prey population, the home range that provides a maximum encounter rate is determined by the overlap between the predator home range and multiple prey home ranges. This predator home range can differ greatly from that which maximizes a given pairwise encounter rate. More importantly, for more crowded situations, the optimal predator home-range size could also be determined by the overlap between the various prey home ranges (prey packing). This suggests the possibility of a tradeoff between maximizing the per-prey-individual encounter rate and maximizing the number of available prey. For example, under what conditions does increasing the predator home range size result in gaining access to enough new prey individuals that the overall prey encounter rate still increases despite decreasing per-prey-individual encounter rates? Scaling our results to more crowded populations will eventually facilitate revisiting the large amount of ecological theory constructed upon the principle of mass action and investigating the conditions in which more realistic encounter models can qualitatively change the outcome of the population models in which they are embedded.

Finally, we have considered here a pair of OU processes, representing a prey and predator that move independently from each other. Even though this is a reasonable first approximation that simplifies the analytical calculations, long-range perception and the use of sensory information by both predator and prey result in attraction-avoidance forces between individuals that may compete with home-range attraction \citep{folmer2010well, Hein2013, Potts2014,Barbier2016}. These long-range interactions introduce cross-correlations between individual trajectories, whose effect on encounter statistics has been treated only through numerical simulations \citep{Martinez-Garcia2013, Martinez-Garcia2014} or very simplified scenarios \citep{Martinez-Garcia2015}. Incorporating them into the analytical framework started here constitutes another direction for future work.
 
\section*{Acknowledgments}

This work is supported by the Gordon \& Betty Moore Foundation through grant GBMF2550.06 to RMG. JMC, CHF, and WFF were supported by award 1458748 from the US NSF Advances in Biological Informatics program. 

%\bibliography{references}

\newpage

\begin{appendices}
\counterwithin{figure}{section}
\numberwithin{equation}{section}

\section{The effect of the encounter kernel} \label{app:encounter-kernel}

\begin{figure}[H]
    \centering
        \includegraphics[width=0.99\textwidth]{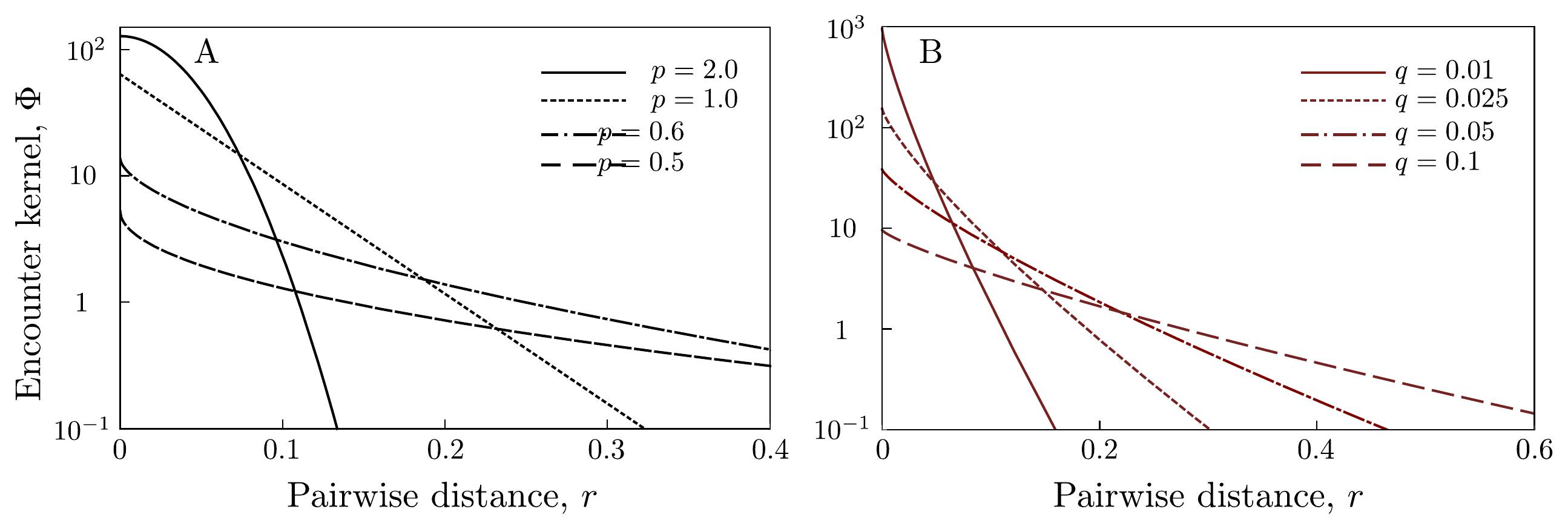}
        \caption{Encounter kernels defined by Eq.~(\ref{eq:genkernels}) for different values of the shape parameter $p$ and $q=0.05$.}\label{fig:kernels}
\end{figure}

In this Appendix, we consider the effect of the shape of the encounter kernel on encounter rates. In particular, we extend the results of the main text considering a family of normalized encounter kernels that are defined by,
\begin{equation}\label{eq:genkernels}
 \Phi_{q}(r) = \frac{\exp\left(-r/q\right)^p}{ \Gamma\left(\frac{2+p}{p} \right) \pi q^2},
\end{equation}
where $q$ gives the spatial scale of perception and $p$ is a positive parameter that controls the steepness of the kernel. In the limit $p\rightarrow\infty$, these kernels converges to the top-hat kernel used in the main text, whereas they show fatter tails as $p$ decreases (Fig.~\ref{fig:kernels}). The gamma function, $\Gamma$, defined as
\begin{equation}
 \Gamma(z) = \int_0^\infty x^{z-1} {\rm e}^{-x} dx
\end{equation}
ensures the normalization of the kernel. 

\begin{figure}[H]
    \centering
        \includegraphics[width=0.9\textwidth]{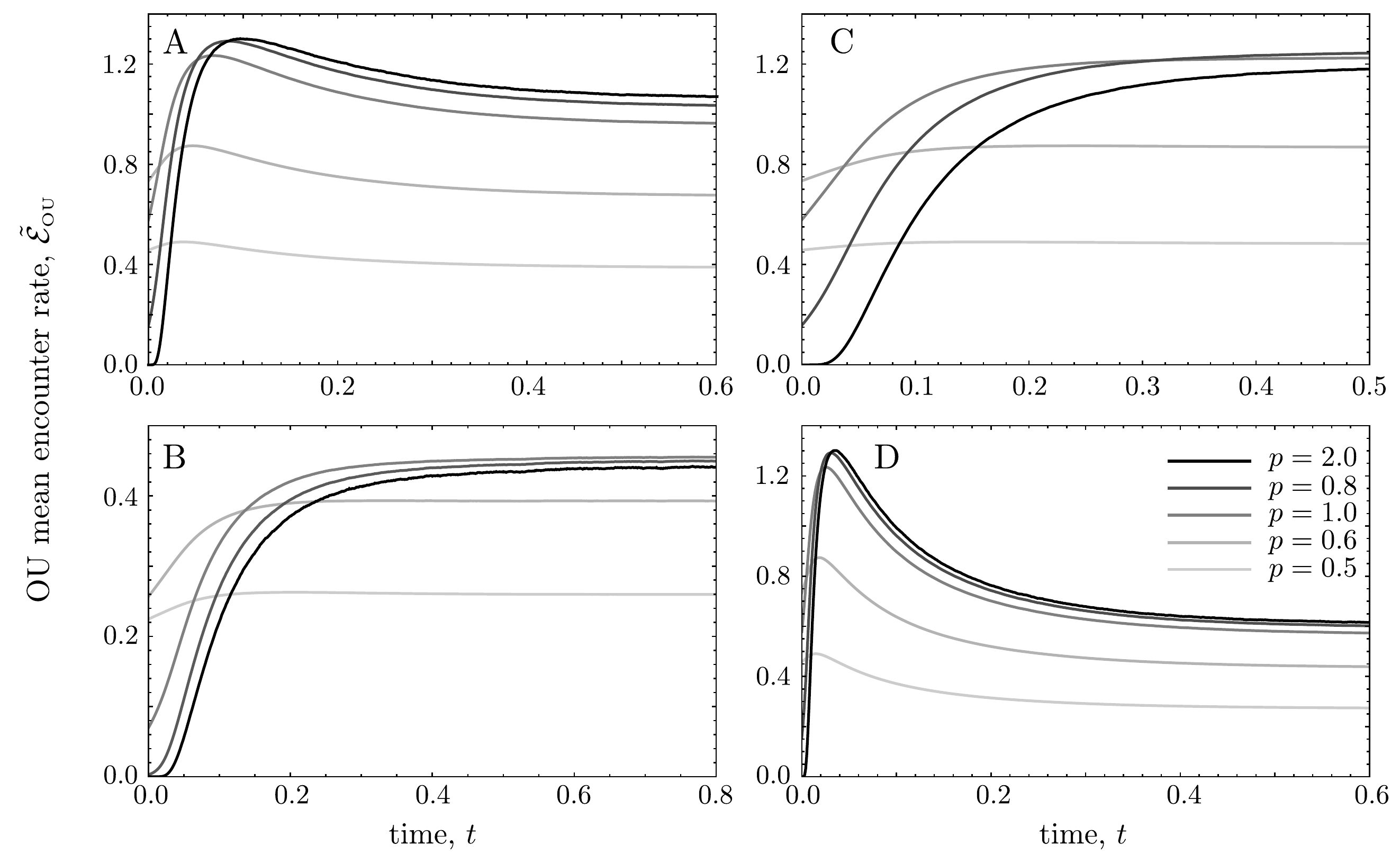}
        \caption{Effect of the encounter kernel on OU mean encounter rates. a) $R_\lambda = 0.3$, $g=0.3$; b) $R_\lambda = 0.3$, $g=0.1$; c) $R_\lambda = 0.5$, $g=0.3$; d) $R_\lambda = 0.3$, $g=0.7$. $\tau = 0.3$ and $q=0.05$ in all the panels.}\label{fig:eff-kernels}
\end{figure}

Our results show a complex relationship between the mean encounter rate and the shape of the encounter kernel controlled by the spatial distribution and size of the home ranges (Fig.~\ref{fig:eff-kernels}). For fatter-tail kernels defined by low $p$, the encounter rate tends to become constant earlier in time (long-dashed lines in Fig.~\ref{fig:eff-kernels}), because the home range of the prey is always within the extension of the encounter kernel. In addition, $p$ controls a short-term vs long-term encounter tradeoff, similar to the one controlled by $g$ and discussed for top-hat kernels in the main text (Fig.~\ref{fig:transient}). Sharper, high-$p$ kernels favor long-term encounters but penalize them in the short-term; smoother, low-$p$ kernels favor short-term encounters at the cost of reduced long-term ones.

\newpage

\section{Brownian motion derivations} \label{app-BM}

Brownian Motion (BM) is obtained in the limit in which OU has vanishing home range affinity, $\tau^{-1}\rightarrow0$, and the movement of the individuals is not bounded. Taking this limit in Eqs.~(\ref{mean-pos}) and (\ref{var-pos}) we obtain,
\begin{eqnarray}
 \mu_\beta(t) &=& \mu_\beta(0), \label{mean-pos-BM} \\
 \sigma^2(t) &=& \sigma^2(0) + gt. \label{var-pos-BM}
\end{eqnarray}

The mean position thus remains equal to its initial value, and because individuals have no home-range affinity, the variance diverges linearly with time (unrestricted movement). In addition, individual space occupation is uniform in the long-time limit. Starting from the same initial condition used to study transient OU encounter rates in the main text, that is, a fixed (deterministic) prey-predator distance equal to $R_\lambda$, the BM noncentrality parameter and rescaling variance, obtained from Eq.~(\ref{mean-pos-BM}) and (\ref{var-pos-BM}), are
\begin{eqnarray}
\Lambda &=& R_\lambda^2 \label{lambdapar-BM} \\
\sigma_r^2(t) &=& 2gt. \label{sigmapar-BM}
\end{eqnarray}

Inserting Eqs.~(\ref{lambdapar-BM}) and (\ref{sigmapar-BM}) into Eq.~(\ref{eq:encOU}) returns the BM mean encounter rate, which tends to zero in the long-time limit. This is a consequence of considering unbounded movement, which makes the variance in the position of both individuals diverge on the long time. At short times, however, RBM and BM provide identical results because the movement statistics in both cases are the same and the effect of the boundary conditions on RBM is still negligible. As border reflections become more important in RBM, BM and RBM encounter rates diverge from each other: $\tilde{\mathcal{E}}_{\mbox{\tiny{RBM}}}$ stabilizes in its stationary value whereas $\tilde{\mathcal{E}}_{\mbox{\tiny{BM}}}$ decays to zero (Fig. \ref{fig:transient}).

\newpage

\section{Moments of the OU model} \label{app-ou}

We start from the description of the OU movement in terms of a stochastic differential equation for the one of the coordinates of the $2D$ position, Eq.~(\ref{ou-pos}),
\begin{equation}
  \dot{z}_\beta(t)= -\frac{1}{\tau}\left[z_\beta(t) - \lambda_\beta \right] + \sqrt{g} \ \xi_\beta(t), \label{ou-sde}\\
\end{equation}
and our goal is to obtain expression for the mean, $\mu_\beta$, and the variance, $\sigma_\beta^2$, of the $2D$ position $(\beta \in \lbrace x,y \rbrace)$. Notice that because $g$ and $\tau$ are the same for each component of $ \bm{\dot{z}}$ they do not have a subscript in Eq.~(\ref{ou-sde}). In the following, we will also remove the subscript $\beta$ from $z_\beta$, $\lambda_\beta$ and $\xi_\beta$ to make the notation simpler. 

First, we integrate Eq.~(\ref{ou-sde}),
  \begin{equation} \label{eq-integral}
  z(t) = z(0){\rm e}^{-t/\tau} + \frac{\lambda}{\tau} \int_0^t ds \ {\rm e}^{-(t-s)/\tau} + g\int_0^t dW(s) \ {\rm e}^{-(t-s)/\tau},
 \end{equation}
where $dW(s) = \xi(s)ds$ is a Wiener process. Solving the deterministic equation, we get 
 \begin{equation}
  z(t) = z(0){\rm e}^{-t/\tau} + \lambda\left[ 1-{\rm e}^{-t/\tau} \right] + g\int_0^t dW(s) \ {\rm e}^{-(t-s)/\tau}, \label{int-two}
 \end{equation}
and finally, taking averages on both sides of Eq.~(\ref{int-two}) the stochastic integral vanishes because $\left\langle dW(s)\right\rangle=0$ and we obtain the mean position in one of the dimensions 
\begin{equation} \label{mpos}
 \mu_z(t) = \mu_z(0) {\rm e}^{-t/\tau} + \lambda\left[ 1-{\rm e}^{-t/\tau} \right],
\end{equation}
where $\mu_z(0) = \left\langle z(0) \right\rangle$ assuming that the initial condition is stochastic.

Next, we calculate the variance of the position. To this end, we start from the definition of the variance, $\sigma_z^2 = \mu_{z^2} - \mu_z^2$ and use Eqs.~(\ref{int-two}) and (\ref{mpos})
to obtain $\mu_{z^2}$ and $\mu_z^2$ respectively,

\begin{eqnarray}
\sigma_z^2(t) &=& \left\langle\left(z(0){\rm e}^{-t/\tau} + \lambda\left[ 1-{\rm e}^{-t/\tau} \right]\right)^2\right\rangle  + g\left\langle\left(\int_0^t dW(s) \ {\rm e}^{-(t-s)/\tau}\right)^2\right\rangle \nonumber \\
	   && + \left\langle2g\left(z(0){\rm e}^{-t/\tau} + \lambda\left[ 1-{\rm e}^{-t/\tau} \right]\right)\int_0^t dW(s) \ {\rm e}^{-(t-s)/\tau}\right\rangle \nonumber \\
	   && - \left( \mu_z(0) {\rm e}^{-t/\tau} + \lambda\left[ 1-{\rm e}^{-t/\tau} \right]\right )^2.
\end{eqnarray}

Because $\left\langle dW(s)\right\rangle = 0$, and calculating all the other mean values,
\begin{equation}\label{ou-varp}
\sigma_z^2(t) = [\mu_{z^2}(0)-\mu_z^2(0)] {\rm e}^{-2t/\tau}  + g\left\langle\left(\int_0^t dW(s) \ {\rm e}^{-(t-s)/\tau}\right)^2\right\rangle
\end{equation}

To calculate the mean value of the square of the integral in Eq.~(\ref{ou-varp}), we use Ito isometry
\begin{equation}\label{itoiso}
 \left\langle\left(\int_0^t dW(s) \ {\rm e}^{-(t-s)/\tau}\right)^2\right\rangle = \left\langle\int_0^t ds \  {\rm e}^{-2(t-s)/\tau}\right\rangle = \frac{\tau}{2} \left(1-{\rm e}^{-2t/\tau}\right)
\end{equation}

Finally, because the first term on the right-hand side of Eq.~(\ref{ou-varp}) is the variance of the initial condition, and substituting the second term by the result of Eq.~(\ref{itoiso}), we obtain
the variance in the position for OU motion with random initial condition,
 \begin{equation}\label{ou-varfinal}
\sigma_z^2(t) =  \sigma_z^2(0) {\rm e}^{-2t/\tau}  + \frac{g\tau}{2} \left(1-{\rm e}^{-2t/\tau}\right),
\end{equation}
which is Eq.~(\ref{var-pos}) in the main text. 
 
\newpage
 
\section{Calculation of the mean encounter rate}\label{app-der}

We depart from the definition of the mean encounter rate in Eq.~(\ref{eq:mean-encounter})
\begin{equation} \label{eq:mean-encounter-app}
 \tilde{\mathcal{E}}(t) = \int \mathcal{E}(r,t) f(r,t)dr,
\end{equation}
which using the piecewise definition of the encounter kernel in Eq.~(\ref{enc-kernel}) becomes,
\begin{equation} \label{eq:appb2}
 \tilde{\mathcal{E}}(t) = \frac{\gamma}{\pi q^2} \int_{0}^{q}  f(r,t)dr.
\end{equation}
Next, we change the variable in Eq.~(\ref{eq:appb2}) from $r$ to $u$ so we can use Eq.~(\ref{pdf-2Ddist}) for the PDF of the nondimensional square distance $u$,
\begin{equation}
 \tilde{\mathcal{E}}(t) = \frac{\gamma}{\pi q^2} \int_{0}^{\left(\frac{q}{\sigma_r}\right)^{2}} f\left(\! u;\tilde{\Lambda}\! \right)du,
\end{equation}
and finally, using the cumulative distribution function for the noncentral $\chi^2$-distribution, we arrive to Eq.~(\ref{eq:encOU}) in the main text
\begin{equation}\label{eq:enc-rate-app}
 \tilde{\mathcal{E}}(t) =  \frac{\gamma}{\pi q^2}\left[ 1 - Q_1\left(\frac{\sqrt{\Lambda}}{\sigma_r},\frac{q}{\sigma_r}\right)\right],
\end{equation}
where $Q_M(a,b)$ is the Marcum-Q-function defined as
\begin{equation}
 Q_M(a,b) = \exp\left( -\frac{a^2 + b^2}{2} \right) \sum_{k=1-M}^{\infty} \left(\frac{a}{b} \right)^{k} I_k(ab),
\end{equation}
where $I_k$ is the Modified Bessel function of the first kind and order $k$. 

\newpage

\section{$\tilde{\mathcal{E}}$ in the small $q$ limit} \label{app:expansion}

The exact expressions for the mean encounter rate, $\tilde{\mathcal{E}}$, involve complicated expressions and special functions, both for OU and RBM movement models. In this Appendix, we derive approximated expressions for the mean encounter rate in the stationary state. In this limit, we can truncate the series expansion of the pairwise distance PDF at the quadratic order, which simplifies its integration and the expressions for the mean encounter rate.

\subsection{Ornstein-Uhlenbeck models}\label{approximation-OU}
We depart from the distribution of pairwise nondimensional distances in the stationary state, Eq.~(\ref{pdf-2Ddist}), 
\begin{equation}\label{pdf-2Ddist-app}
 f_{\mbox{\tiny{OU}}}(u;\tilde{\Lambda}) = \frac{1}{2}\exp\left(\frac{-(u+\tilde{\Lambda})}{2}\right) I_0\left(\! \sqrt{\tilde{\Lambda} u} \right).
\end{equation}

First, we obtain the Taylor series of $f_{\mbox{\tiny{OU}}}$ up to linear order. Notice that because $u\propto r^2$, our approximation is up to quadratic order in the pairwise distance,
\begin{eqnarray}
 \exp\left(-\frac{u}{2}\right) &=& 1 - \frac{1}{2}u + \mathcal{O}(u^2), \label{series-exp} \\
 I_0\left(\sqrt{\tilde{\Lambda}u}\right) &=& 1 + \frac{\tilde{\Lambda}}{4}u + \mathcal{O}(u^2) \label{series-bessel}
\end{eqnarray}

Multiplying the expansions in Eq.~(\ref{series-exp}) and (\ref{series-bessel}) and rearranging the additional terms in Eq.~(\ref{pdf-2Ddist-app}), we obtain the linear approximation to Eq.~(\ref{pdf-2Ddist-app}),
\begin{equation}
 f_{\mbox{\tiny{OU}}}(u;\tilde{\Lambda}) = \frac{1}{2}\exp\left(-\frac{\tilde{\Lambda}}{2}\right)\left[ 1 - \frac{u}{2}\left(1-\frac{\tilde{\Lambda}}{2}\right)\right] + \mathcal{O}(u^2).
\end{equation}

Next, to calculate the mean encounter rate, we must integrate the pairwise distance distribution,
\begin{eqnarray}
   \tilde{\mathcal{E}}_{\mbox{\tiny{OU}}}(t\rightarrow\infty,q) &\approx& \frac{\gamma}{\pi q^2}\int_0^{\left(\frac{q}{\sigma_r}\right)^2}f_{\mbox{\tiny{OU}}}(u;\tilde{\Lambda}) du \nonumber \\
  &=& \frac{\gamma}{\pi q^2\sigma_r^2} \int_0^q f_{\mbox{\tiny{OU}}}(r;\Lambda) r dr \nonumber \\
  &=& \frac{\gamma \exp\left(-\frac{\Lambda}{2\sigma_r^2}\right)}{16\pi\sigma_r^6}\left[8\sigma_r^4 + q^2(\Lambda - 2\sigma_r^2) \right].
\end{eqnarray}

\subsection{Reflected Brownian Motion} \label{approximation-BM}
We depart from Eq.~(\ref{bm-dist}), that gives the stationary RBM pairwise distance PDF when $0 \leq r \leq 2\mathcal{R}$,
\begin{equation}\label{bm-dist-app}
 f_{\mbox{\tiny{RBM}}}(r,t\rightarrow\infty) =                                          
\frac{4r}{\pi \mathcal{R}^2}{\rm ArcCos}\left(\frac{r}{2\mathcal{R}}\right) - \frac{2r^2}{\pi \mathcal{R}^4}\sqrt{\mathcal{R}^2 - \frac{r^2}{4}}
\end{equation}
in the limit $r<<\mathcal{R}$, we can truncate the Taylor series for $f_{\mbox{\tiny{RBM}}}$ at the quadratic order
\begin{equation}
 f_{\mbox{\tiny{RBM}}}(r,t\rightarrow\infty) = \frac{2r}{\mathcal{R}^2} - \frac{4r^2}{\pi \mathcal{R}^3} + \mathcal{O}(r^4),
\end{equation}
and integrate it to obtain the mean instantaneous encounter rate
\begin{equation}\label{bm-enc-app}
 \tilde{\mathcal{E}}_{\mbox{\tiny{RBM}}}(t\rightarrow\infty,q)  \approx  \frac{\gamma}{\pi\mathcal{R}^2}\left( 1 - \frac{4 q}{3\pi \mathcal{R}}\right).
\end{equation}

\newpage

\section{OU encounter rate versus home range overlap}\label{app:overlap}

In this Appendix, we present the calculation of the overlap between two home ranges using both the Bhattacharyya coefficient (BC) and the scalar product of the individual position PDFs, and write the pairwise encounter rate in terms of it. We consider two OU processes with, in general, time-dependent mean, $\bm{\mu}_i(t)$, and covariance matrix, $\bm{\Sigma}_i(t)$, defined by the mean and the variance of Eqs.~(\ref{mean-pos}) and (\ref{var-pos}) in the main text,
 \begin{equation}\label{eq:OU1}
 \begin{array}{cc}
\bm{\mu}_j(t) = \left( \begin{array}{c} \mu_{j,x}(t) \\ \mu_{j,y}(t)  \end{array}\right) & \bm{\Sigma}_j(t) = \left( \begin{array}{cc} \sigma_j^2(t) & 0 \\ 0 & \sigma_j^2(t)  \end{array}\right)\\
\end{array}
\end{equation}
where $j=1,2$ for the predator and the prey respectively and we have already considered that individual movement is isotropic. Therefore, the $x$-$y$ covariances are zero and the covariance matrices are scalar matrices.

\subsection{The Bhattacharyya coefficient (BC)}
For two continuous probability distributions $f(x)$ and $g(x)$, it is defined as
\begin{equation}
 BC(f,g) = \int \sqrt{f(x)g(x)}dx.
\end{equation}

For the particular case in which $f$ and $g$ are bivariate Normal distributions, a closed expression for the BC can be derived \citep{Abou-Moustafa2010},
\begin{equation}\label{eq:BC}
 BC = \frac{\rvert \bm{\Sigma}_1 \rvert ^{1/4}\rvert \bm{\Sigma}_2 \rvert ^{1/4}}{\rvert \bm{\Sigma} \rvert^{1/2}} \exp\left(-\frac{1}{8}\bm{\mu}^T \bm{\Sigma}^{-1} \bm{\mu} \right),
\end{equation}
where $\bm{\Sigma} = (\bm{\Sigma}_1 + \bm{\Sigma}_2)/2$ and $\bm{\mu} = \bm{\mu}_1 - \bm{\mu}_2$. Inserting the mean and the covariance matrices given by Eq.~(\ref{eq:OU1}) in Eq.~(\ref{eq:BC}), we obtain
\begin{equation}\label{BC:alpha}
 BC(t) = 2 \frac{\sigma_1(t)\sigma_2(t)}{\sigma_r^2(t)}\exp\left(-\frac{\Lambda(t)}{4\sigma_r^2(t)}\right).
\end{equation}

On the other hand,  the OU encounter rate is
\begin{equation}\label{eq:appOUen}
\tilde{\mathcal{E}}_{\mbox{\tiny{OU}}}(t, q=0) = \frac{\gamma}{2\pi \sigma_r^2(t)} \exp\left(-\frac{\Lambda(t)}{2 \sigma_r^2(t)}\right),
\end{equation}
where we have taken $q=0$ in Eq.~(\ref{eq:app-encOU}) to simplify the calculations.

Next, from Eq.~(\ref{BC:alpha}), we obtain
\begin{equation}
 \exp\left(-\frac{\Lambda}{2\sigma_r^2}\right) = \frac{\sigma_r^4}{4 (\sigma_1 \sigma_2)^2}BC(t)^2
\end{equation}
that can be inserted in Eq.~(\ref{eq:appOUen}) to obtain the OU encounter rate as a function of the home range overlap, BC,
\begin{equation}\label{eq:appOUenBC}
\tilde{\mathcal{E}}_{\mbox{\tiny{OU}}}(t, q=0) = \frac{\gamma \sigma_r^2(t)}{8 \pi \sigma_1^2(t) \sigma_2^2(t)}BC(t)^2
\end{equation}
Eq.~(\ref{eq:appOUenBC}) reveals a quadratic scaling law between encounter rate and home-range overlap with the encounter rate at maximum overlap, $BC=1$, depending on home range areas through $\sigma_1^2$ and $\sigma_2^2$.

\subsection{PDFs scalar product}
Alternatively, we can quantify the overlap between home ranges using the scalar product of the individual position PDFs. Given two function, $f(x)$ and $g(x)$, we define the scalar product, $f \!\cdot\! g$, as
\begin{equation}
f\!\cdot\! g = \int f(x) g(x) dx.
\end{equation}

For the particular case in which $f$ and $g$ are the individual position PDFs, namely $f_1$ and $f_2$, we have to compute the scalar product of two bivariate Normal distributions with mean and covariance matrices defined by Eqs.~(\ref{eq:OU1}). We obtain
\begin{equation}
f_1(x,y;t)\! \cdot\! f_2(x,y;t) = \int \! dx dy \  f_1(x,y;t) f_2(x,y;t) = \frac{1}{2\pi \sigma^2_r(t)} \exp\left(-\frac{\Lambda(t)}{2\sigma_r^2(t)}\right),
\end{equation}
and hence
\begin{equation}\label{eq:appOUenBC-V2}
\tilde{\mathcal{E}}_{\mbox{\tiny{OU}}}(t, q=0) = \gamma \ \ f_1(x,y;t)\!\cdot\!f_2(x,y;t).
\end{equation}
\newpage

\section{Effect of predator home range.}\label{app:eff-homerange}

In this Appendix, we provide detailed results on the effect that predator home ranges have on OU stationary encounter rates. To simplify the calculations, we consider the local-perception limit ($q=0$) of Eq.~(\ref{eq:encOU}) and to simplify the notation we do not write explicitly the stationary-state condition $t\rightarrow\infty$. With this considerations, the stationary OU mean encounter rate is,
\begin{equation}\label{eq:app-encOU-app}
   \tilde{\mathcal{E}}_{\mbox{\tiny{OU}}}(q=0) = \frac{\gamma}{2\pi\sigma_r^2} \exp\left(-\frac{\Lambda}{2\sigma_r^2}\right).
\end{equation}
where the predator home range radius enters through the definition of the rescaling variance, $\sigma_r^2$. 

\begin{figure}[H]
  \centering
   \includegraphics[width=0.55\textwidth]{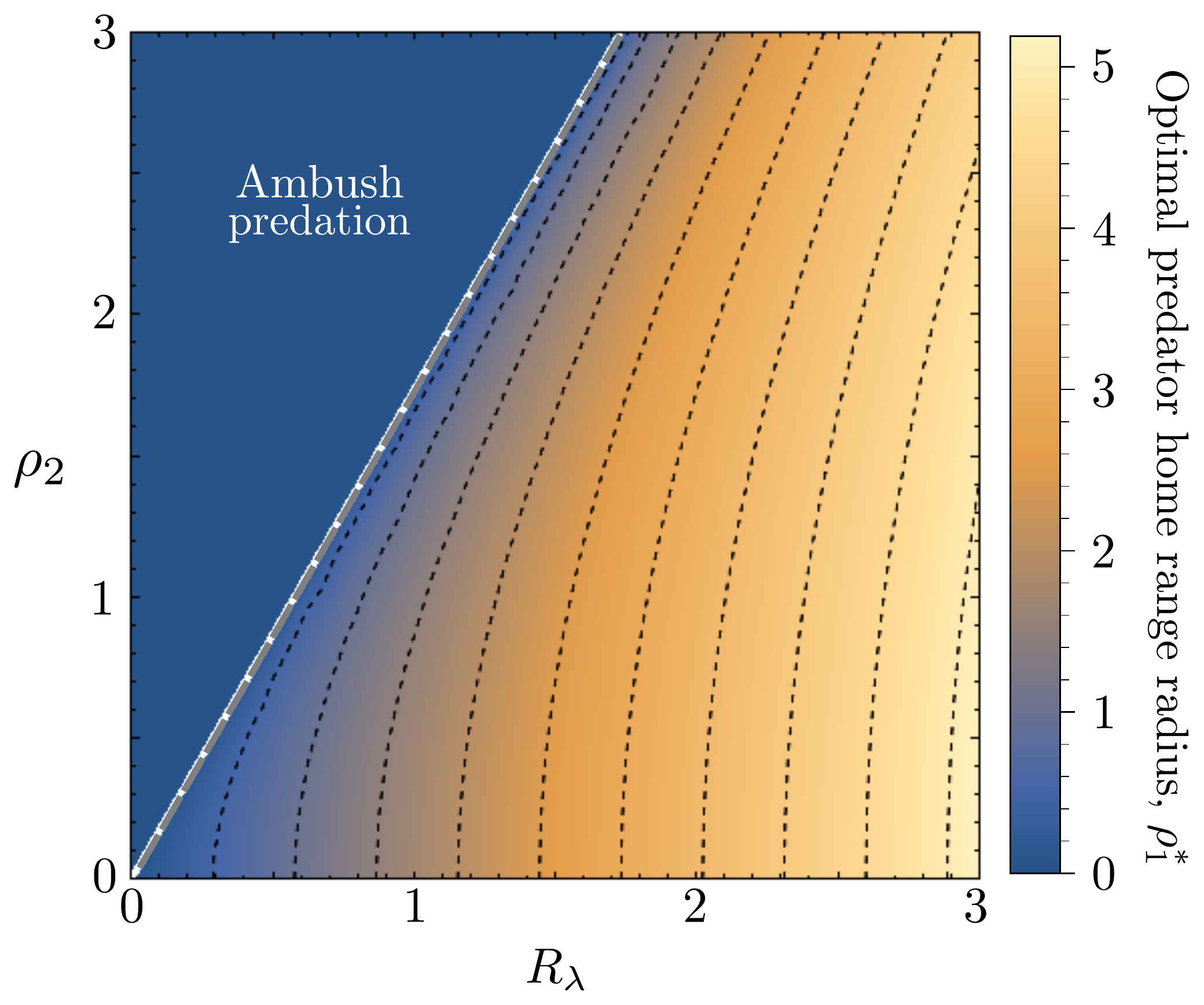}
   \caption{OU optimal predator home range radius $\rho_1^{*}$ for different habitat structures, represented by the distance between home-range centers, $R_\lambda$, and prey home range radii, $\rho_2$. The white dashed line traces the transition from $\rho_1^*=0$ (ambush predation) to some exploration ($\rho_1^{*} \neq 0$).}
    \label{fig:shortq-app}
\end{figure}

First, we calculate the optimal predator home range size, $\rho_1^*$, that maximizes the encounter rate for a given prey home range radius, $\rho_2$, and distance between home range centers, $R_\lambda$. This optimal predator home range size indicates how much space the predator should explore, depending on the configuration of the landscape and the movement of the prey, to maximize its predation rate. Because we are interested in the $\rho_1$ that maximizes the encounter rate, we need to solve, 
\begin{equation}
\frac{\partial \tilde{\mathcal{E}}_{\mbox{\tiny{OU}}}}{\partial \rho_1} = 0,
\end{equation}
for $\rho_1$. Performing the derivatives, we obtain
\begin{equation}\label{eq:dereps-app}
 \frac{\gamma}{2\pi}\left(\frac{\Lambda - 2\sigma_r^2}{\sigma_r^6}\right) \exp\left(-\frac{\Lambda}{2\sigma_r^2}\right)\frac{\rho_1}{K^2} = 0,
\end{equation}
which has two solutions. The first one, $\rho_1^* = 0$, accounts for the existence of a maximum or local minimum of the encounter rate for ambush predation. The second one is obtained from solving $\Lambda - 2\sigma_r^2$, which gives, using the definition of the noncentrality parameter and the rescaling variance in Eqs.~(\ref{lambdapar-st}) and (\ref{sigmapar-st}),
\begin{equation}\label{eq:optpredator}
\rho_1^* = \sqrt{K^2R_\lambda^2 - \rho_2^2}.
\end{equation}
Because the home range radius must be positive, Eq.~(\ref{eq:optpredator}) indicates that the encounter rate is maximum for ambushing strategies if $R_\lambda < \rho_2/K$. This condition is met if the prey frequently crosses the predator home-range center, which may happen either if $R_\lambda$ is small (home-ranges are close to each other) or if $\rho_2$ is large (the prey shows a high mobility and therefore a large home range). If neither of these conditions are met, the mean encounter rate is maximum for some predator mobility given by Eq.~(\ref{eq:optpredator}) (Figure \ref{fig:shortq-app}). 

In addition, using the short-perception expansion of the encounter rate, 
\begin{equation}\label{eq:app2-encOU}
   \tilde{\mathcal{E}}_{\mbox{\tiny{OU}}}(q) \approx \frac{\gamma \left[8\sigma_r^4 + q^2(\Lambda - 2\sigma_r^2) \right]}{16\pi\sigma_r^6} \exp\left(-\frac{\Lambda}{2\sigma_r^2}\right).
\end{equation}
we can derive scaling relationships between the encounter rate and the predator home range radius when all the other model parameters are kept constant.  For instance, in nomadic predators, represented by large home ranges ($\rho_1\rightarrow\infty$ and hence $\sigma_r^2\rightarrow\infty$) the encounter rate is a monotonically decreasing function of the predator home-range size, decaying as $\rho_1^{-2}$. This result follows from the fact that the exponential term in Eq.~(\ref{eq:app2-encOU}) tends to a constant when $\sigma_r\rightarrow\infty$ and the dominant terms in the numerator and the denominator of Eq.~(\ref{eq:app2-encOU}) are of fourth and sixth order in $\rho_1$ respectively. The opposite limit, $\rho_1 \rightarrow 0$, represents ambushing strategies by sedentary predators that do not abandon their home-range centers. The encounter rate in this limit can be obtained by retaining up to the linear terms in $\rho_1$ on an additional Taylor expansion of Eq.~(\ref{eq:app2-encOU}) around $\rho_1=0$. 

\newpage

\section{Steady-state encounter rates for Reflected Brownian Motion}\label{app:rbm-steady}
In this appendix, we calculate the stationary mean encounter rate between a predator-prey pair of RBM individuals that move within the population ranges defined by the set of OU movement parameters used in the main text to study the steady-state OU mean encounter rate. To establish a more direct comparison with OU results, we will present RBM results as a function of the underlying OU movement parameters that define the population range. In particular, using Eq.~(\ref{radius}), we obtain the radius of the population range, $\mathcal{R}$, as a function of the distance between home-range centers, $R_\lambda$, and the size of the individual OU home ranges, $\rho_1$ and $\rho_2$.

\begin{figure}[H]
    \centering
        \includegraphics[width=\textwidth]{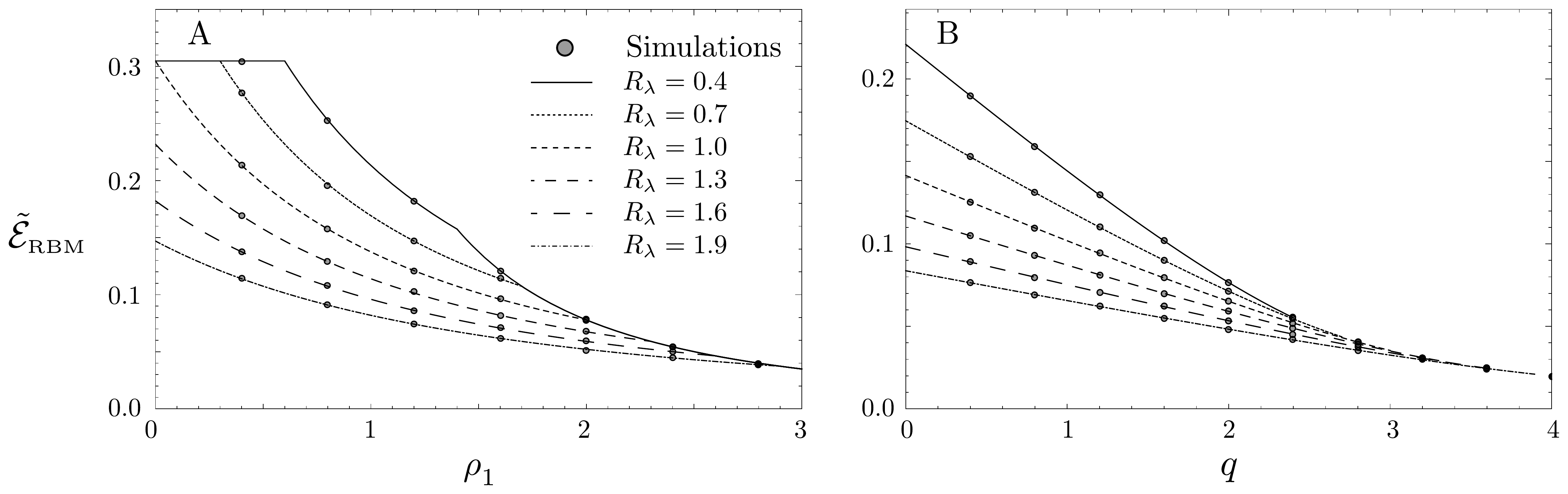}
        \caption{RBM mean encounter rate versus $\rho_1$ with $q=0.1$ (A) and versus $q$ with $\rho_1 = 1$ (B). Symbols correspond to numerical simulations. Changes in the slope of the curves in panel (A) are due to jumps in $\mathcal{R}$ given by Eq.~(\ref{radius}).}
        \label{fig:aier-bbm}
\end{figure}

 We find that the mean encounter rate first remains constant for small $\rho_1$ and then decreases as $\rho_1$ increases, while it is a monotonically decreasing function of $q$. Because RBM individuals visit all the regions of the encounter arena with the same frequency and they do not have home range affinity, changing either $\rho_1$ or $R_\lambda$ only affects $\mathcal{R}$, without influencing the overlap between individual position PDFs, which is always maximum. Therefore, larger $\rho_1$ and $R_\lambda$ may lead to larger $\mathcal{R}$ according, and because encounters occur in a larger arena they are less likely (smaller encounter rate). In fact, considering the definition of the population range, if $\rho_1<\rho_2-R_\lambda$, then $\mathcal{R}=\rho_2$ and the encounter rate is constant (solid line in Fig. \ref{fig:aier-bbm}A). For larger $\rho_1$, the size of the encounter arena increases with $\rho_1$ and hence the encounter rate decreases (Fig.~\ref{fig:aier-bbm}A). How quickly the mean encounter rate decreases with $\rho_1$ depends on the branch of Eq.~(\ref{radius}) that determines the radius of the encounter arena. Moreover, Eq.~(\ref{bm-enc-app-main}) recovers the linear decay of the encounter rate with $q$ at short perception ranges (Fig.~\ref{fig:aier-bbm}B). This is an important difference between OU and RBM encounters; while nonlocal perception ($q > 0$) maximizes the OU mean encounter rate when home ranges are small and far from each other (Fig.~\ref{fig:aier-bmou}B), the RBM mean encounter rate is always maximum for $q=0$ (Fig.~\ref{fig:aier-bbm}B; Appendix \ref{app:effq} for further details). 

\newpage

\section{Effect of the perceptual range.}\label{app:effq}

In this Appendix, we extend the calculations for the effect of nonlocal predator perception on encounter rates. We depart from the steady state limit ($t\rightarrow\infty$) of the OU mean encounter rate in the short-perception limit, Eq.~(\ref{eq:app-encOU}) in the main text, 
\begin{equation}\label{appendix-short}
   \tilde{\mathcal{E}}_{\mbox{\tiny{OU}}}(q) \approx \frac{\gamma \left[8\sigma_r^4 + q^2(\Lambda - 2\sigma_r^2) \right]}{16\pi\sigma_r^6} \exp\left(-\frac{\Lambda}{2\sigma_r^2}\right).
\end{equation}

\begin{figure}[H]
    \centering
        \includegraphics[width=\textwidth]{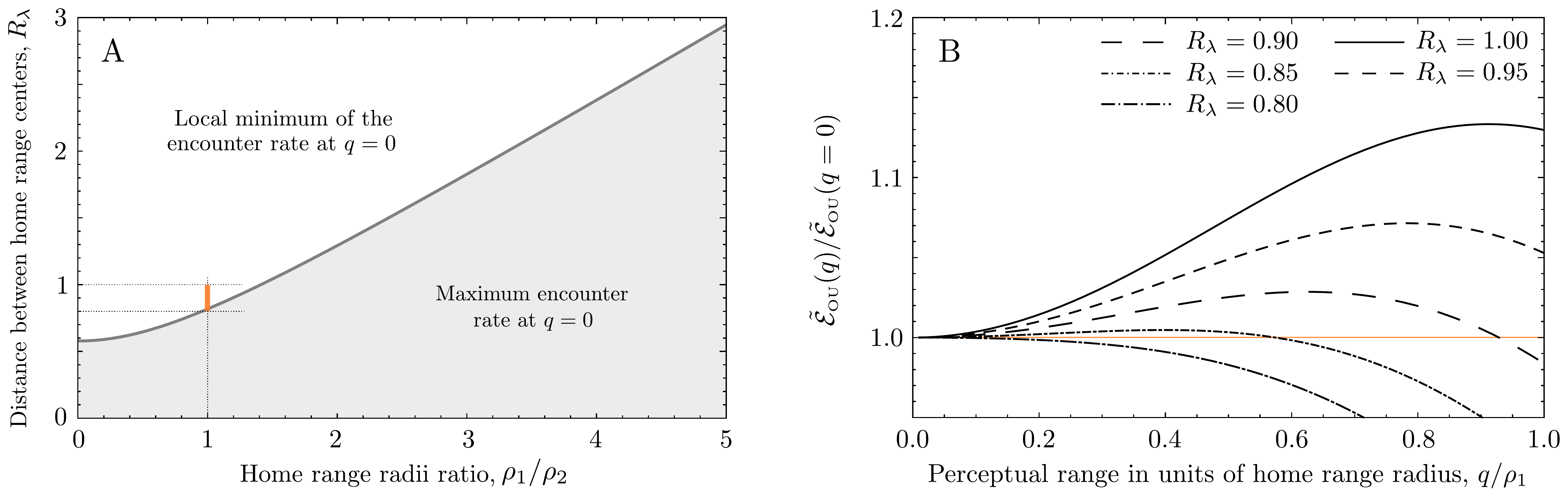}
        \caption{A) Phase diagram for the benefits of long-range versus local perception depending on the distance between home-range centers and their relative sizes. In the gray region, the OU mean encounter rates is maximum for local perception ($q=0$).  In the white region, $q=0$ is a local minimum of the encounter rate and there exists an optimal intermediate perceptual range. The orange rectangle indicates the subregion of the parameter space covered in panel B. B) Ratio between the OU mean encounter range and its value for local perception, $q=0$, as a function of the predator perceptual range normalized by home range radius. As the distance between predator and prey home range centers increases, the optimal perceptual range increases, and it eventually reaches unrealistic values $q>\rho_1$.}
        \label{fig:nonlocal}
\end{figure}

From its second derivative with respect to $q$, it follows that the mean encounter rate is maximum for local perception ($q=0$) if $\Lambda < 2\sigma_r^2$. If $\Lambda > 2\sigma_r^2$, however, $q=0$ is a local minimum and hence the encounter rate increases if the perceptual range increases. Using nondimensional quantities, we can write the condition for this switch in the behavior of the encounter rate as (Fig.~\ref{fig:nonlocal}A),
\begin{equation}
 \frac{R_\lambda}{\rho_2} > \sqrt{1+\left(\frac{\rho_1}{\rho_2}\right)^2}.
\end{equation}

Interestingly, in the limit $q\rightarrow\infty$, $\tilde{\mathcal{E}}_{\mbox{\tiny{OU}}} \sim q^{-2}$, which means that the mean encounter rate tends to zero as the perceptual range becomes infinitely large. Therefore, when the OU mean encounter rate, $\tilde{\mathcal{E}}_{\mbox{\tiny{OU}}}$, has a local minimum at $q=0$, an intermediate range of perception $q > 0$ may maximize the encounter rate. The existence of such optimal mid-range perception aligns with results from our own previous studies in which we investigated the interplay between long-range information gathering and foraging efficiency in different contexts \citep{Martinez-Garcia2013, Martinez-Garcia2014, Martinez-Garcia2017, Fagan2017,colombo2019spatial}. Because Eq.~(\ref{appendix-short}) gives an approximated OU mean encounter rate that is only accurate in the short-perception limit, we cannot, in general, obtain an analytical expression for the optimal perceptual range when it is larger than zero. Instead, we need to evaluate the exact expression in terms of the Marcum-Q-function, Eq.~(\ref{eq:encOU}) in the main text,
\begin{equation}\label{eq:encOU-app}
  \tilde{\mathcal{E}}(q) = \frac{\gamma}{\pi q^2}\left[ 1 - Q_1\left(\frac{\sqrt{\Lambda}}{\sigma_r},\frac{q}{\sigma_r}\right)\right].
\end{equation}

This analysis reveals that, when optimal perception is achieved for $q>0$, the optimal perceptual range scales quickly with $R_\lambda$ and can lead to an unrealistic regime in which the optimal perceptual range exceeds the radius of the predator home range (Fig.~\ref{fig:nonlocal}B).

\newpage

\section{Supplementary figures}

\begin{figure}[H]
    \centering
        \includegraphics[width=0.95\textwidth]{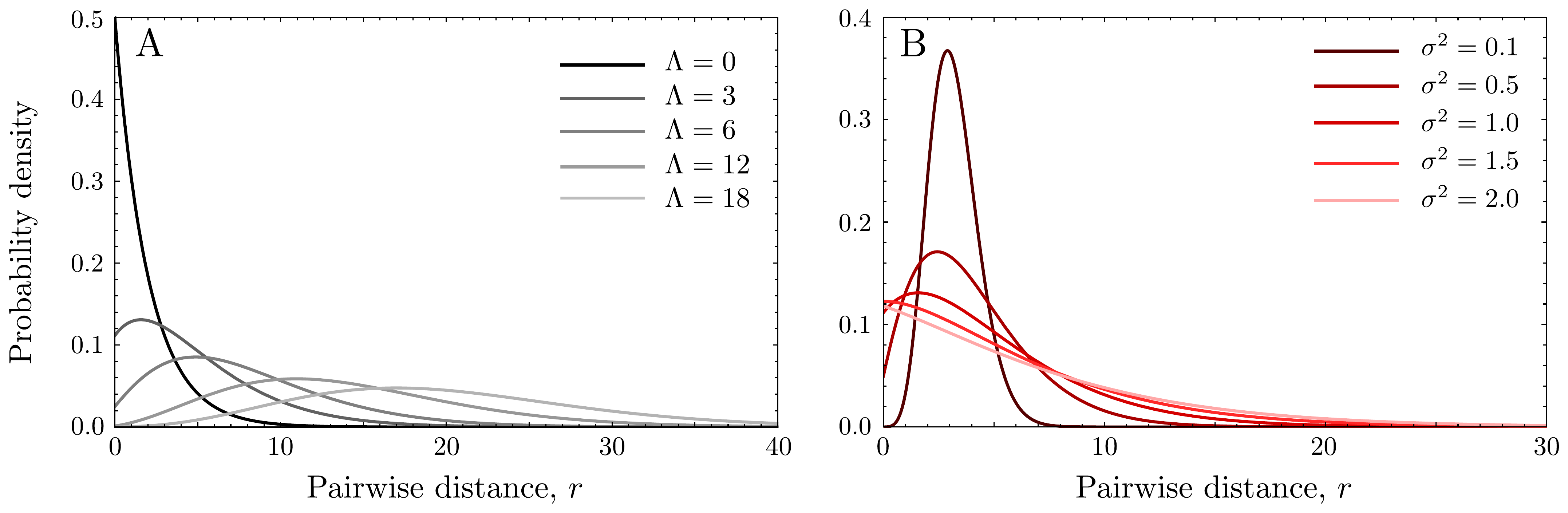}
        \caption{Probability density function of the noncentral chi-squared distribution. A)Effect of the noncentrality parameter, $\Lambda$ with fixed $\sigma = 1$. B) Effect of $\sigma$ with fixed $\Lambda=3$.}\label{fig:distpdf}
\end{figure}

\begin{figure}[H]
     \centering
        \includegraphics[width=0.49\textwidth]{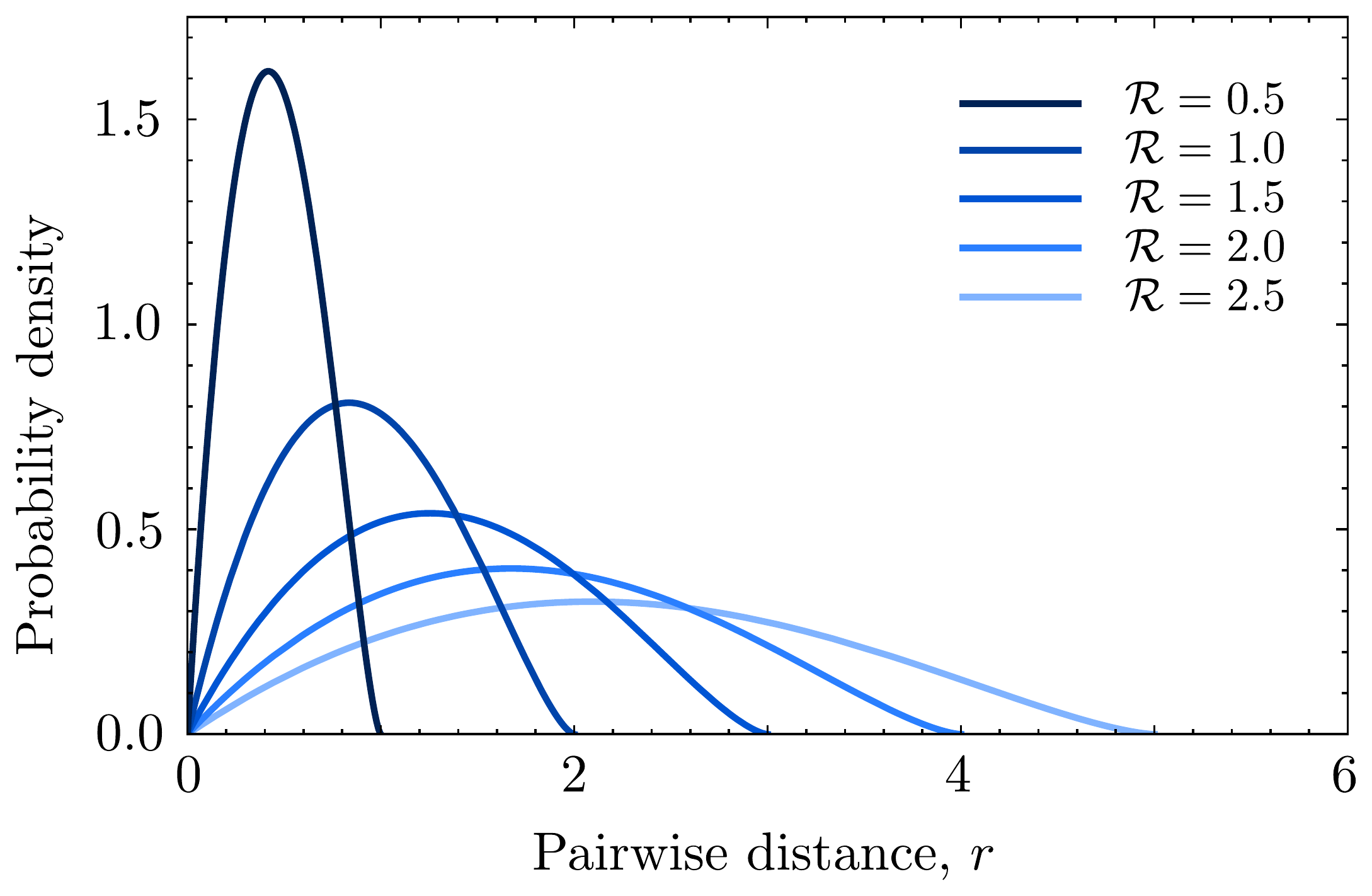}
        \caption{Stationary pairwise distance distribution for RBM. Different curves correspond to different population ranges, $\mathcal{R}$.}\label{fig:distpdfbm}
\end{figure}

\begin{figure}[H]
    \centering
        \includegraphics[width=0.75\textwidth]{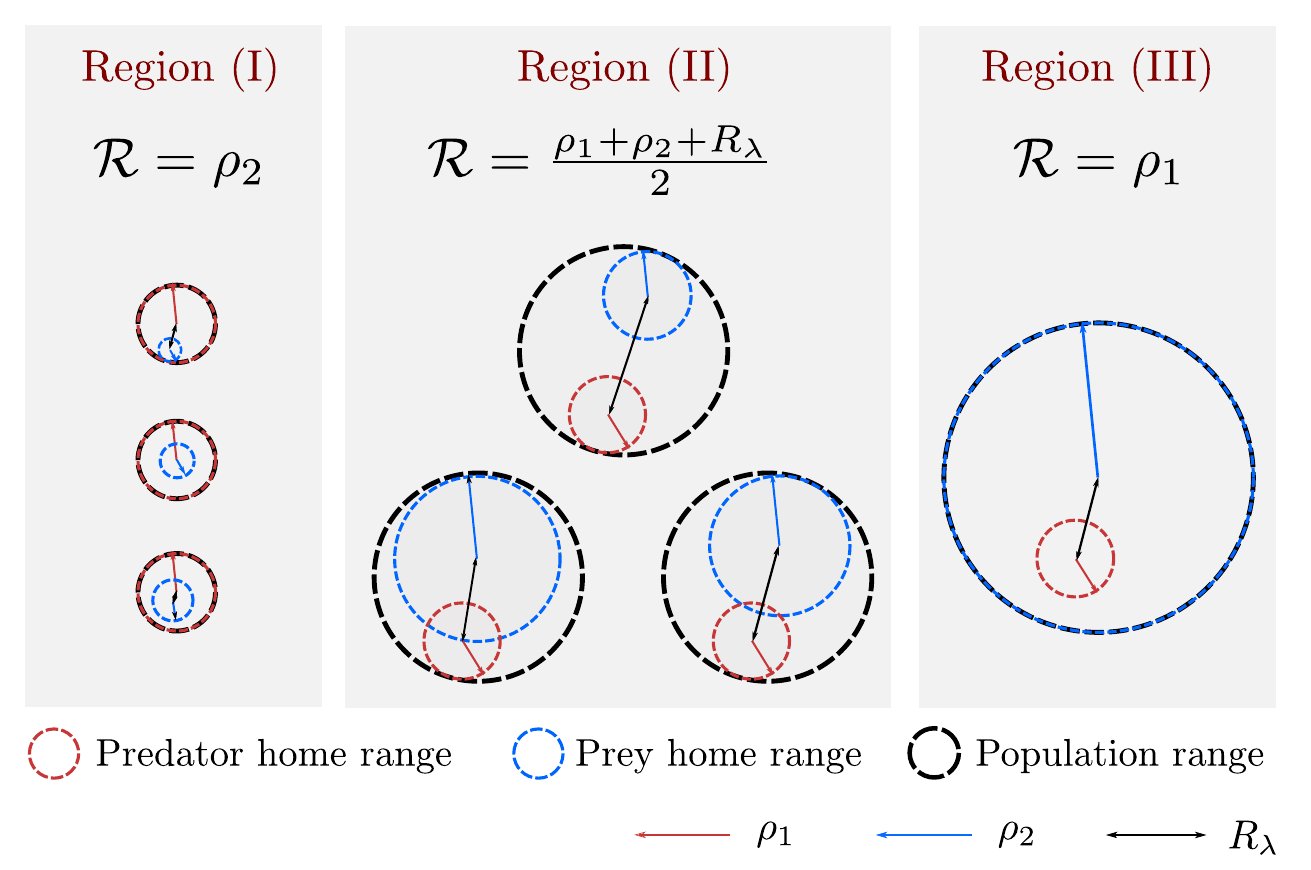}
        \caption{Schematic representation of how individual home ranges are distributed within the population range for different parameterizations of the OU models. Region (I), (II) and (III) are defined according to the branch of Eq.~(\ref{radius}) that determines the population range, that is, they depend on the radius of the prey and the predator home range and the distance between their centers.}
        \label{fig:scheme}
\end{figure}

\begin{figure}[H]
    \centering
        \includegraphics[width=0.97\textwidth]{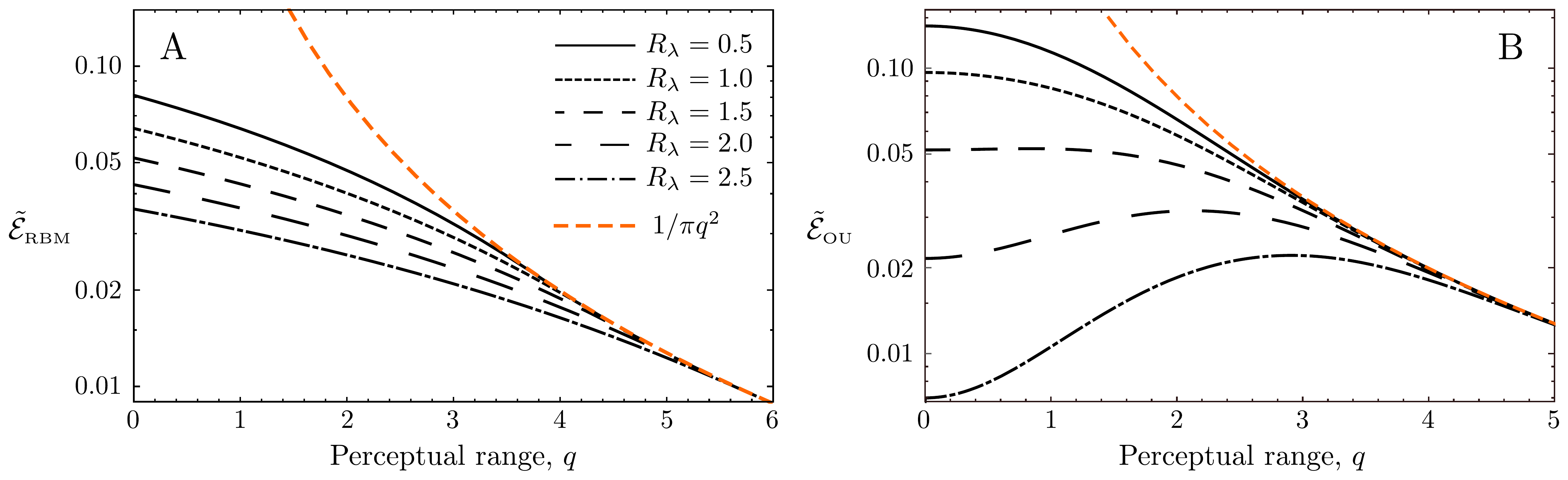}
        \caption{Log-linear plot of the encounter rates in the stationary state for RBM (A) and OU (B) models. In the limit $q\rightarrow\infty$ the encounter rates decay as $q^{-2}$.}\label{fig:encasymptotic}
\end{figure}

\end{appendices}

\end{document}